\documentclass[aps, prc, reprint, superscriptaddress, amsmath, amssymb]{revtex4-1}
\usepackage[T1]{fontenc}
\usepackage{graphicx}
\usepackage{dcolumn}
\usepackage{bm}
\usepackage[version=3]{mhchem}
\usepackage{xcolor}
\usepackage{xspace}
\usepackage[normalem]{ulem}
\usepackage[acronym,toc,shortcuts]{glossaries}
\usepackage{multirow}
\usepackage{hyperref}
\makeglossaries
\newacronym{DCS}{DCS}{differential cross section}

\begin{document}

\newcommand{\dSdOexp}{$\left(\frac{d\sigma}{d\Omega}\right)_{\mathrm{exp}}$}
\newcommand{\dSdOthry}{$\left(\frac{d\sigma}{d\Omega}\right)_{\mathrm{th}}$}

\newcommand{\ZeroM}{$0^{-}$\xspace}
\newcommand{\OneM}{$1^{-}$\xspace}
\newcommand{\TwoM}{$2^{-}$\xspace}
\newcommand{\ThreeM}{$3^{-}$\xspace}

\newcommand{\piSD}{\ce{\pi (sd)}\xspace}
\newcommand{\piSOneDFour}{\ce{\pi s^{\,\,\,\,1}_{1/2} d^{\,\,\,\,4}_{3/2}}\xspace}
\newcommand{\piSTwoDThree}{\ce{\pi s^{\,\,\,\,2}_{1/2} d^{\,\,\,\,3}_{3/2}}\xspace}

\newcommand{\nuFP}{\ce{\nu (fp)}\xspace}
\newcommand{\nuPonehalf}{\ce{\nu p_{1/2}}\xspace}
\newcommand{\nuPthreehalf}{\ce{\nu p_{3/2}}\xspace}
\newcommand{\nuFfivehalf}{\ce{\nu f_{5/2}}\xspace}
\newcommand{\nuFsevenhalf}{\ce{\nu f_{7/2}}\xspace}

\newcommand{\Ponehalf}{\ce{p_{1/2}}\xspace}
\newcommand{\Pthreehalf}{\ce{p_{3/2}}\xspace}
\newcommand{\Ffivehalf}{\ce{f_{5/2}}\xspace}
\newcommand{\Fsevenhalf}{\ce{f_{7/2}}\xspace}

\newcommand{\piShole}{\ce{\pi s^{-1}_{1/2}}\xspace}
\newcommand{\piDhole}{\ce{\pi d^{-1}_{3/2}}\xspace}

\hyphenation{VAMOS}
\hyphenation{MUGAST}
\hyphenation{AGATA}

\title{Direct transfer to $^{46,48}$K as a survey of the $\pi(s_{1/2})$$-\nu(sdpf)$ interaction}
\author{C.\,J.~Paxman}
    \affiliation{School of Maths and Physics, University of Surrey, Guildford, GU2 7XH, United Kingdom}
    \affiliation{Grand Acc\'{e}l\'{e}rateur National d’Ions Lourds (GANIL), CEA/DRF-CNRS/IN2P3, Bvd Henri Becquerel, 14076 Caen, France}
\author{A.~Matta}
    \affiliation{Université de Caen Normandie, ENSICAEN, CNRS/IN2P3, LPC Caen UMR6534, F-14000 Caen, France}
\author{W.\,N.~Catford}
    \affiliation{School of Maths and Physics, University of Surrey, Guildford, GU2 7XH, United Kingdom}
\author{G.~Lotay}
    \affiliation{School of Maths and Physics, University of Surrey, Guildford, GU2 7XH, United Kingdom}
\author{M.~Assi\'e}
    \affiliation{Universit\'{e} Paris-Saclay, CNRS/IN2P3, IJCLab, 91405 Orsay, France}
\author{E.~Clément}
    \affiliation{Grand Acc\'{e}l\'{e}rateur National d’Ions Lourds (GANIL), CEA/DRF-CNRS/IN2P3, Bvd Henri Becquerel, 14076 Caen, France}
\author{A.~Lemasson}
    \affiliation{Grand Acc\'{e}l\'{e}rateur National d’Ions Lourds (GANIL), CEA/DRF-CNRS/IN2P3, Bvd Henri Becquerel, 14076 Caen, France}
\author{D.~Ramos}
    \affiliation{Grand Acc\'{e}l\'{e}rateur National d’Ions Lourds (GANIL), CEA/DRF-CNRS/IN2P3, Bvd Henri Becquerel, 14076 Caen, France}
\author{N.\,A.~Orr}
    \affiliation{Université de Caen Normandie, ENSICAEN, CNRS/IN2P3, LPC Caen UMR6534, F-14000 Caen, France}
\author{F.~Galtarossa}
    \affiliation{Universit\'{e} Paris-Saclay, CNRS/IN2P3, IJCLab, 91405 Orsay, France}
    \affiliation{INFN Sezione di Padova, I-35131 Padova, Italy}
    \affiliation{Dipartimento di Fisica e Astronomia dell'Università di Padova, I-35131 Padova, Italy}
\author{V.~Girard-Alcindor}
    \affiliation{Grand Acc\'{e}l\'{e}rateur National d’Ions Lourds (GANIL), CEA/DRF-CNRS/IN2P3, Bvd Henri Becquerel, 14076 Caen, France}
    \affiliation{Universit\'{e} Paris-Saclay, CNRS/IN2P3, IJCLab, 91405 Orsay, France}
\author{J.~Dudouet}
    \affiliation{Universit\'{e} Claude Bernard Lyon 1, CNRS/IN2P3, IP2I Lyon, UMR 5822, F-69100 Villeurbanne, France}
\author{N.\,L.~Achouri}
    \affiliation{Université de Caen Normandie, ENSICAEN, CNRS/IN2P3, LPC Caen UMR6534, F-14000 Caen, France}
\author{D.~Ackermann}
    \affiliation{Grand Acc\'{e}l\'{e}rateur National d’Ions Lourds (GANIL), CEA/DRF-CNRS/IN2P3, Bvd Henri Becquerel, 14076 Caen, France}
\author{D.~Barrientos}
    \affiliation{CERN, CH-1211 Geneva 23 (Switzerland)}
\author{D.~Beaumel}
    \affiliation{Universit\'{e} Paris-Saclay, CNRS/IN2P3, IJCLab, 91405 Orsay, France}
\author{P.~Bednarczyk}
    \affiliation{The Henryk Niewodniczański Institute of Nuclear Physics, Polish Academy of Sciences, ul. Radzikowskiego 152, 31-342 Kraków, Poland}
\author{G.~Benzoni}
    \affiliation{INFN Sezione di Milano, I-20133 Milano, Italy}
\author{A.~Bracco}
    \affiliation{INFN Sezione di Milano, I-20133 Milano, Italy}
    \affiliation{Dipartimento di Fisica, Università di Milano, I-20133 Milano, Italy}
\author{L.~Canete}
    \affiliation{School of Maths and Physics, University of Surrey, Guildford, GU2 7XH, United Kingdom}
\author{B.~Cederwall}
    \affiliation{Department of Physics, KTH Royal Institute of Technology, SE-10691 Stockholm, Sweden}
\author{M.~Ciemala}
    \affiliation{The Henryk Niewodniczański Institute of Nuclear Physics, Polish Academy of Sciences, ul. Radzikowskiego 152, 31-342 Kraków, Poland}
\author{P.~Delahaye}
    \affiliation{Grand Acc\'{e}l\'{e}rateur National d’Ions Lourds (GANIL), CEA/DRF-CNRS/IN2P3, Bvd Henri Becquerel, 14076 Caen, France}
\author{D.\,T.~Doherty}
    \affiliation{School of Maths and Physics, University of Surrey, Guildford, GU2 7XH, United Kingdom}
\author{C.~Domingo-Pardo}
    \affiliation{Instituto de Física Corpuscular, CSIC-Universidad de Valencia, E-46071 Valencia, Spain}
\author{B.~Fern\'andez-Dom\'inguez}
    \affiliation{IGFAE and Dpt. de F\'{i}sica de Part\'{i}culas, Univ. of Santiago de Compostela, E-15758, Santiago de Compostela, Spain}
\author{D.~Fern\'andez} 
    \affiliation{IGFAE and Dpt. de F\'{i}sica de Part\'{i}culas, Univ. of Santiago de Compostela, E-15758, Santiago de Compostela, Spain}
\author{F.~Flavigny}
    \affiliation{Université de Caen Normandie, ENSICAEN, CNRS/IN2P3, LPC Caen UMR6534, F-14000 Caen, France}
\author{C.~Foug\`{e}res}
    \affiliation{Grand Acc\'{e}l\'{e}rateur National d’Ions Lourds (GANIL), CEA/DRF-CNRS/IN2P3, Bvd Henri Becquerel, 14076 Caen, France}
\author{G.~de~France}
    \affiliation{Grand Acc\'{e}l\'{e}rateur National d’Ions Lourds (GANIL), CEA/DRF-CNRS/IN2P3, Bvd Henri Becquerel, 14076 Caen, France}
\author{S.~Franchoo}
    \affiliation{Universit\'{e} Paris-Saclay, CNRS/IN2P3, IJCLab, 91405 Orsay, France}
\author{A.~Gadea}
    \affiliation{Instituto de Física Corpuscular, CSIC-Universidad de Valencia, E-46071 Valencia, Spain}
\author{J.~Gibelin}
    \affiliation{Université de Caen Normandie, ENSICAEN, CNRS/IN2P3, LPC Caen UMR6534, F-14000 Caen, France}
\author{V.~González}
    \affiliation{Departamento de Ingeniería Electrónica, Universitat de Valencia, Burjassot, Valencia, Spain}
\author{A.~Gottardo}
    \affiliation{Laboratori Nazionali di Legnaro, INFN, I-35020 Legnaro (PD), Italy}
\author{N.~Goyal}
    \affiliation{Grand Acc\'{e}l\'{e}rateur National d’Ions Lourds (GANIL), CEA/DRF-CNRS/IN2P3, Bvd Henri Becquerel, 14076 Caen, France}
\author{F.~Hammache}
    \affiliation{Universit\'{e} Paris-Saclay, CNRS/IN2P3, IJCLab, 91405 Orsay, France}
\author{L.\,J.~Harkness-Brennan}
    \affiliation{Oliver Lodge Laboratory, The University of Liverpool, Liverpool, L69 7ZE, UK}
\author{D.\,S.~Harrouz}
    \affiliation{Universit\'{e} Paris-Saclay, CNRS/IN2P3, IJCLab, 91405 Orsay, France}
\author{B.~Jacquot}
    \affiliation{Grand Acc\'{e}l\'{e}rateur National d’Ions Lourds (GANIL), CEA/DRF-CNRS/IN2P3, Bvd Henri Becquerel, 14076 Caen, France}
\author{D.\,S.~Judson}
    \affiliation{Oliver Lodge Laboratory, The University of Liverpool, Liverpool, L69 7ZE, UK}
\author{A.~Jungclaus}
    \affiliation{Instituto de Estructura de la Materia, CSIC, Madrid, E-28006 Madrid, Spain}
\author{A.~Kaşkaş}
    \affiliation{Department of Physics, Faculty of Science, Ankara University, 06100 Besevler - Ankara, Turkey}
\author{W.~Korten}
    \affiliation{Irfu, CEA, Université Paris-Saclay, F-91191 Gif-sur-Yvette, France}
\author{M.~Labiche}
    \affiliation{STFC Daresbury Laboratory, Daresbury, Warrington, WA4 4AD, UK}
\author{L.~Lalanne}
    \affiliation{Grand Acc\'{e}l\'{e}rateur National d’Ions Lourds (GANIL), CEA/DRF-CNRS/IN2P3, Bvd Henri Becquerel, 14076 Caen, France}
    \affiliation{Universit\'{e} Paris-Saclay, CNRS/IN2P3, IJCLab, 91405 Orsay, France}
\author{C.~Lenain}
    \affiliation{Université de Caen Normandie, ENSICAEN, CNRS/IN2P3, LPC Caen UMR6534, F-14000 Caen, France}
\author{S.~Leoni}
    \affiliation{INFN Sezione di Milano, I-20133 Milano, Italy}
    \affiliation{Dipartimento di Fisica, Università di Milano, I-20133 Milano, Italy}
\author{J.~Ljungvall}
    \affiliation{Universit\'{e} Paris-Saclay, CNRS/IN2P3, IJCLab, 91405 Orsay, France}
\author{J.~Lois\,-Fuentes}
    \affiliation{IGFAE and Dpt. de F\'{i}sica de Part\'{i}culas, Univ. of Santiago de Compostela, E-15758, Santiago de Compostela, Spain}
\author{T.~Lokotko}
    \affiliation{Université de Caen Normandie, ENSICAEN, CNRS/IN2P3, LPC Caen UMR6534, F-14000 Caen, France}
\author{A.~Lopez-Martens}
    \affiliation{Universit\'{e} Paris-Saclay, CNRS/IN2P3, IJCLab, 91405 Orsay, France}  
\author{A.~Maj}
    \affiliation{The Henryk Niewodniczański Institute of Nuclear Physics, Polish Academy of Sciences, ul. Radzikowskiego 152, 31-342 Kraków, Poland}
\author{F.\,M.~Marqu\'{e}s}
    \affiliation{Université de Caen Normandie, ENSICAEN, CNRS/IN2P3, LPC Caen UMR6534, F-14000 Caen, France}
\author{I.~Martel}
    \affiliation{Departamento de Ciencias Integradas, Universidad de Huelva, Calle Dr.~Cantero Cuadrado, 6, 21004 Huelva, Spain}
\author{R.~Menegazzo}
    \affiliation{INFN Sezione di Padova, I-35131 Padova, Italy}
\author{D.~Mengoni}
    \affiliation{INFN Sezione di Padova, I-35131 Padova, Italy}
    \affiliation{Dipartimento di Fisica e Astronomia dell'Università di Padova, I-35131 Padova, Italy}
\author{B.~Million}
    \affiliation{INFN Sezione di Milano, I-20133 Milano, Italy}
\author{J.~Nyberg}
    \affiliation{Department of Physics and Astronomy, Uppsala University, SE-75120 Uppsala, Sweden}
\author{R.\,M.~Pérez-Vidal}
    \affiliation{Instituto de Física Corpuscular, CSIC-Universidad de Valencia, E-46071 Valencia, Spain}
    \affiliation{Laboratori Nazionali di Legnaro, INFN, I-35020 Legnaro (PD), Italy}
\author{L.~Plagnol}
    \affiliation{Université de Caen Normandie, ENSICAEN, CNRS/IN2P3, LPC Caen UMR6534, F-14000 Caen, France}
\author{Zs.~Podolyák}
    \affiliation{School of Maths and Physics, University of Surrey, Guildford, GU2 7XH, United Kingdom}
\author{A.~Pullia}
    \affiliation{INFN Sezione di Milano, I-20133 Milano, Italy}
    \affiliation{Dipartimento di Fisica, Università di Milano, I-20133 Milano, Italy}
\author{B.~Quintana}
    \affiliation{Laboratorio de Radiaciones Ionizantes, Departamento de Física Fundamental, Universidad de Salamanca, E-37008 Salamanca, Spain}
\author{D.~Regueira-Castro}
    \affiliation{IGFAE and Dpt. de F\'{i}sica de Part\'{i}culas, Univ. of Santiago de Compostela, E-15758, Santiago de Compostela, Spain}
\author{P.~Reiter}
    \affiliation{Institut für Kernphysik, Universität zu Köln, Zülpicher Str. 77, D-50937 Köln, Germany}
\author{M.~Rejmund}
    \affiliation{Grand Acc\'{e}l\'{e}rateur National d’Ions Lourds (GANIL), CEA/DRF-CNRS/IN2P3, Bvd Henri Becquerel, 14076 Caen, France}
\author{K.~Rezynkina}
    \affiliation{Université de Strasbourg, CNRS, IPHC UMR 7178, F-67000 Strasbourg, France}
    \affiliation{INFN Sezione di Padova, I-35131 Padova, Italy}
\author{E.~Sanchis}
    \affiliation{Departamento de Ingeniería Electrónica, Universitat de Valencia, Burjassot, Valencia, Spain}
\author{M.~Şenyiğit}
    \affiliation{Department of Physics, Faculty of Science, Ankara University, 06100 Besevler - Ankara, Turkey}
\author{N.~de~S\'er\'eville}
    \affiliation{Universit\'{e} Paris-Saclay, CNRS/IN2P3, IJCLab, 91405 Orsay, France}
\author{M.~Siciliano}
    \affiliation{Laboratori Nazionali di Legnaro, INFN, I-35020 Legnaro (PD), Italy}
    \affiliation{Irfu, CEA, Université Paris-Saclay, F-91191 Gif-sur-Yvette, France}
    \affiliation{Physics Division, Argonne National Laboratory, Lemont (IL), United States}
\author{D.~Sohler} 
    \affiliation{Institute for Nuclear Research, Atomki, 4001 Debrecen, P.O. Box 51, Hungary}
\author{O.~Stezowski}
    \affiliation{Universit\'{e} Claude Bernard Lyon 1, CNRS/IN2P3, IP2I Lyon, UMR 5822, F-69100 Villeurbanne, France}
\author{J.-C.~Thomas}
    \affiliation{Grand Acc\'{e}l\'{e}rateur National d’Ions Lourds (GANIL), CEA/DRF-CNRS/IN2P3, Bvd Henri Becquerel, 14076 Caen, France}
\author{A.~Utepov}
    \affiliation{Grand Acc\'{e}l\'{e}rateur National d’Ions Lourds (GANIL), CEA/DRF-CNRS/IN2P3, Bvd Henri Becquerel, 14076 Caen, France}
\author{J.\,J.~Valiente-Dobón}
    \affiliation{Laboratori Nazionali di Legnaro, INFN, I-35020 Legnaro (PD), Italy}
\author{D.~Verney}
    \affiliation{Universit\'{e} Paris-Saclay, CNRS/IN2P3, IJCLab, 91405 Orsay, France}
\author{M.~Zielińska}
    \affiliation{Irfu, CEA, Université Paris-Saclay, F-91191 Gif-sur-Yvette, France}

\date{\today}

\begin{abstract}

The collapse of the canonical $N=28$ magic number in nuclei with $Z<20$ has drawn significant interest as it relates to the emergence of an island of inversion  centered on $^{42}$Si and $^{44}$S. In particular, interactions between the $\pi s_{1/2}$ orbital -- empty in $^{42}$Si and full in $^{44}$S -- and the neutron orbitals just above and below the $N=28$ gap are expected to be critical in this region, but remain relatively unexplored. In this paper, we expand upon the results of our previous study of the direct transfer reaction $^{47}$K(d,p$\gamma$)$^{48}$K [C.\,J.~Paxman \textit{et al.}, Phys. Rev. Lett. 134, 162504 (2025)] with the results of the complementary $^{47}$K(d,t$\gamma$)$^{46}$K reaction. Through this study, we present a comprehensive scan of the interaction between the critical $\pi s_{1/2}$ orbital and a broad range of neutron orbitals spanning nearly two full shells. We identify several discrepancies between the experimental results and state-of-the-art shell model calculations, which suggest a deficiency of the shell model to fully capture the complex proton configuration mixing in this region, highlighting a significant challenge for single-particle descriptions of the island of inversion.

\end{abstract}

\maketitle


\clearpage
\newpage
\mbox{}
\clearpage
\newpage

\section{Introduction}

The nuclear shell model~\cite{Mayer1949_ShellsInNuclei,HaxelJensenSuess1949_SpinOrbSplit,BrownWildenthal1988_ReviewShellModel,Brown2022_ShellModelTowardsDripLines} remains one of the most versatile frameworks for describing the structure of the atomic nucleus, treating it as a quantum system of protons and neutrons occupying discrete orbitals, organized into shells. The energetic gaps between these shells give rise to the well-known “magic numbers” of nucleons, which correspond to particularly stable configurations. However, these shell gaps are known to evolve when moving away from stability due to the shifting balance of interactions between 
orbitals~\cite{Otsuka2005_ShellEvolution}. In order to predict the properties of exotic nuclei, these orbital interactions are characterized -- either phenomenologically or from bare nucleon-nucleon interactions -- in the shell model, and verified by comparison to experimental results. As such, a fundamental goal of experimental studies of nuclear structure is to perform key observations to constrain the behaviors of these myriad orbital interactions. While some interactions are generally well-understood, such as those easily accessible in stable nuclei, we must map out and understand exotic regions of the nuclear chart, where the imbalance of $N$ and $Z$ provides insight into new, unexplored interactions. Refining shell-model interactions in these regions enhances their predictive power, and provides greater insight into the microscopic origins of emergent collective behaviors, such as those found within the $N=28$ island of inversion, where the expected shell closure weakens and increased collectivity is observed, as exemplified by $^{42}$Si~\cite{Grevy2004_CollectivityIn42Si,42Si_N28Collapse_E2energy,gade2019_structure42Si} and $^{44}$S~\cite{Glasmacher1997_44S-Collective,44S_N28Collapse_E2energy,Caceres2012_44S-shapecoex}.

\begin{figure}
    \centering
    \includegraphics[width=0.8\linewidth]{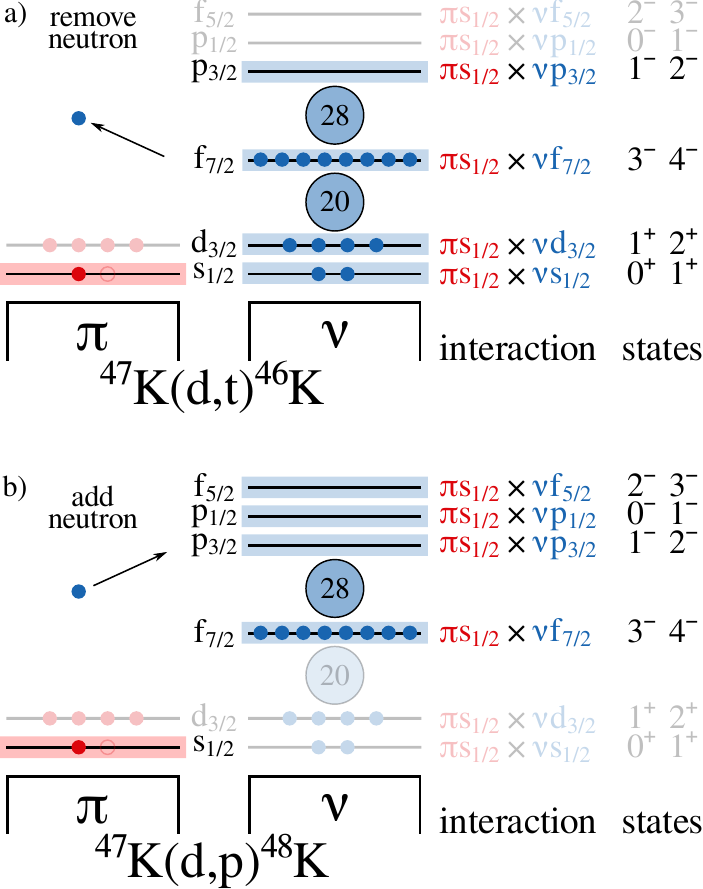}
    \caption{Simplified schematic of the proton (neutron) orbitals and their occupation, shown in red (blue). As the ground state of $^{47}$K is dominated by $\pi(s_{1/2}^1 d_{3/2}^4)$, all states populated in $^{46}$K (a) and $^{48}$K (b) will be accessed via this component. The spin-parities of states that can be produced by coupling this proton to each of the accessible neutron orbitals is shown. Notably, $\nu p_{3/2}$ and $\nu f_{7/2}$ are accessible in both reactions due to the diffuse neutron occupation probabilities across $N=28$.}
    \label{fig:schematic}
\end{figure}

A point of focus, in this regard, relates to understanding nuclear systems where the proton and/or neutron occupancies are markedly changed by the addition or removal of a single nucleon. In particular, the nucleus $^{47}$K is expected to be of special importance; while the dominant ground-state proton configuration of $^{46,48}$K is $\pi(s_{1/2}^{2} d_{3/2}^{3})$, the N=28 isotope $^{47}$K is instead dominated by $\pi(s_{1/2}^1 d_{3/2}^4)$ configuration in the ground state~\cite{Papuga2014}. Our recent study of the $^{47}$K(d,p$\gamma$)$^{48}$K direct transfer reaction~\cite{Paxman2025_47Kdp} established the excited structure of $^{48}$K as a key benchmark for nuclear structure studies involving cross-shell $\pi(sd)$$-\nu(fp)$ interactions, which are critical in the $N=28$ island of inversion. In this article, we present results of the simultaneous $^{47}$K(d,t$\gamma$)$^{46}$K study -- the first study of $^{46}$K by single-nucleon removal. This further contextualizes our understanding of this region by probing neutron orbital \textit{occupation}, looking at the interaction of a single neutron hole in $N=28$ with a single proton in $\pi s_{1/2}$. Additional details of weak, complex and unbound states in $^{48}$K will also be presented. Through the combination of $^{47}$K(d,p) and $^{47}$K(d,t) reactions, this work provides an extensive overview of the interactions between $\pi s_{1/2}$ and $\nu(s_{1/2}, d_{3/2}, f_{7/2}, p_{3/2}, p_{1/2}, f_{5/2})$, spanning two major shells (see Fig.~\ref{fig:schematic}). By exploring neutron configurations situated both above and below the $N=28$ shell gap, and their coupling to the key $\pi s_{1/2}$ proton orbital -- which is unoccupied in $^{42}$Si and fully occupied in $^{44}$S -- this work is a benchmark for predictions of single-particle structure in the $N=28$ island of inversion.

\section{Experimental procedure}

The neutron pick-up and stripping reactions $^{47}$K(d,p)$^{48}$K and $^{47}$K(d,t)$^{46}$K were studied as part of a campaign of experiments~\cite{Assie2021_MAV} coupling the $\gamma$-ray tracking array AGATA~\cite{Akkoyun2012_AgataNIM} and the particle detector array MUGAST~\cite{Assie2021_MAV} to the magnetic spectrometer VAMOS++~\cite{Rejmund2011_VAMOS++}, at the \textit{Grand Accélérateur National d'Ions Lourds} (GANIL) in Caen, France. A radioactive beam of $^{47}$K was provided by the SPIRAL$1$ facility~\cite{Kamalou2019_SPIRAL1+Upgrade,Chauveau2023_SPIRAL1upgrades} using the method of isotope separation on-line. A primary beam of stable $^{48}$Ca bombarded a thick graphite target at 60 MeV/nucleon. The resulting fragments underwent ionisation~\cite{Chauveau2016_GANIL-FEBIAD} and charge breeding~\cite{Delahaye2012_Spiral1Prospects_ChargeBreeding,Chauveau2023_SPIRAL1upgrades}, and this cocktail beam was reaccelerated with the CIME cyclotron, isolating the $^{47}$K$^{10+}$ fragments with a mass resolution of $10^{-4}$, sufficient to ensure $>99.99\%$ beam purity. This radioactive isotope beam was delivered to the experimental hall with an energy of $7.7$~MeV/nucleon and an average intensity of $5\times10^5$~pps, where it impinged on a $0.31(2)$~mg/cm$^{2}$ self-supporting CD$_{2}$ target (4\% by number H to D). The beam was monitored just before the target position with a single CATS~\cite{OttiniHustache1999_CATS} multi-wire proportional counter.

Four reactions were measured simultaneously; the $^{47}$K(d,p) and $^{47}$K(d,t) transfer reactions, $^{47}$K(d,d) elastic scattering and some $^{47}$K(p,p) elastic scattering from the small amount of proton contamination in the target. These reactions were observed through the combined use of the MUGAST, AGATA and VAMOS++ detection systems~\cite{MugastAgataVamos_NIMA}. 

\paragraph{VAMOS++} 
Heavy recoil nuclei were detected in the focal plane of the VAMOS++ magnetic spectrometer~\cite{Rejmund2011_VAMOS++}, positioned at zero degrees relative to the beam line in a fast-counting arrangement. Due to the small mass difference between the beam nucleus and recoil nucleus, the VAMOS++ focal plane was receiving the whole beam; as such, only the fast-counting multi-wire parallel plate avalanche counter (MWPPAC) was employed, and no heavy recoil particle identification was performed. Critically, the B$\rho$ selection inherent in a recoil reaching the focal plane ensures that only $^{47}$K$+d$ or $^{47}$K$+p$ reactions produce a signal in the focal plane detectors, eliminating heavier reaction products such as those from $^{47}$K$+^{12}$C. As such, the timing signal from the MWPPAC is sufficient to reject fusion-evaporation reactions on carbon in the CD$_{2}$ target -- the largest source of background in experiments of this type -- isolating the direct reactions of interest and resulting in spectra with minimal background. 

\paragraph{MUGAST} 
Light ejectile particles were detected in the MUGAST silicon array~\cite{MugastAgataVamos_NIMA}, a transitional phase of the GRIT~\cite{GRITwebsite} project. MUGAST is comprised of; four square MUST2~\cite{Pollacco2005_MUST2TechnicalPaper} double-sided silicon strip detectors (DSSDs) backed with CsI scintillators positioned at forward angles ($\theta_{lab}=4^{\circ}$ to $22^{\circ}$) to observe tritons from the $^{47}$K(d,t) reaction; six trapezoidal GRIT DSSDs at backwards angles ($\theta_{lab}=104^{\circ}$ to $156^{\circ}$) to observe protons from the $^{47}$K(d,p) reaction; and one square MUST2 DSSD placed just forwards of $90^{\circ}$ to detect elastically scattered light ejectiles. The downstream triton spectroscopy required additional background removal procedures; as such, the CsI backings of the DSSD detectors were used as a veto to reject high-energy background, and energy versus time-of-flight particle identification was performed using both the MUGAST-CATS and MUGAST-VAMOS++ timing~\cite{Paxman2024_Thesis}. MUGAST was calibrated using a standard triple-$\alpha$ source consisting of $^{239}$Pu, $^{241}$Am and $^{244}$Cm. The precise geometry of the array was determined post-experiment using a portable six-axis arm. The MUGAST-VAMOS++ timing signal and the angle of the emitted ejectile allow for unambiguous kinematic selection of the reaction channel, and hence, only the observed angle and energy of the light ejectile are required for kinematic reconstruction of the excitation energy of the recoil nucleus.

\paragraph{AGATA} 
The $\gamma$-ray tracking array AGATA~\cite{Akkoyun2012_AgataNIM} was used for the detection of coincident $\gamma$-ray emissions from excited states in $^{46,48}$K. Twelve AGATA triple clusters (36 segmented HPGe crystals) were mounted in the upstream direction, 18~cm from the target \cite{MugastAgataVamos_NIMA}. Energy and efficiency calibrations were performed using a standard $^{152}$Eu source, extrapolated to high energies through simulation~\cite{Farnea2010_AgataSimulations,Clement2017_Agata1piGanil}. Add-back and pulse shape analysis were performed on-line, and Doppler correction was performed using velocities calculated event-by-event; that is to say, the average beam velocity calculated at the midpoint of the target was additionally corrected for the recoiling effect of the transfer reaction using the detected energy and angle of the light ejectile. Due to this precise reconstruction, $\gamma$-rays of energies up to $\sim$4~MeV were observable, and the $\gamma$-ray FWHM was 7~keV at 1.8~MeV. Such high-resolution $\gamma$-ray spectroscopy is critical for determining the precise energies of populated states and constructing decay schemes. 

\paragraph{Data acquisition and analysis} 
The data acquisition system is described in detail in Ref.~\cite{Assie2021_MAV}. The AGATA detector was operated in triggerless mode, whilst VAMOS++, CATS and MUGAST were in common dead time triggered by MUGAST. During this experiment, the VAMOS++, CATS and MUGAST data flow used a event-number-based merger, which was then coupled to AGATA by means of time stamp using the GTS system \cite{Akkoyun2012_AgataNIM} with a 1~$\mu$s event building window. Data analysis was performed using the ROOT data analysis framework (v.\,6.22/02)~\cite{Brun1997_ROOT,ROOT_version} with the nptool analysis and simulation framework (v.\,3)~\cite{Matta_2016_nptool}.

\paragraph{Target thickness and beam spot position}

The reconstruction of the heavy ejectile excitation energy is sensitive to the energy loss of the ejectile moving through the target, requiring the total target thickness to be known very precisely. Target thickness was determined through two complementary methods --  analysis of the elastic scattering data, and numerical minimization of the reconstructed excitation energy of $^{48}$K.

\begin{figure}
    \centering
    \includegraphics[width=\linewidth]{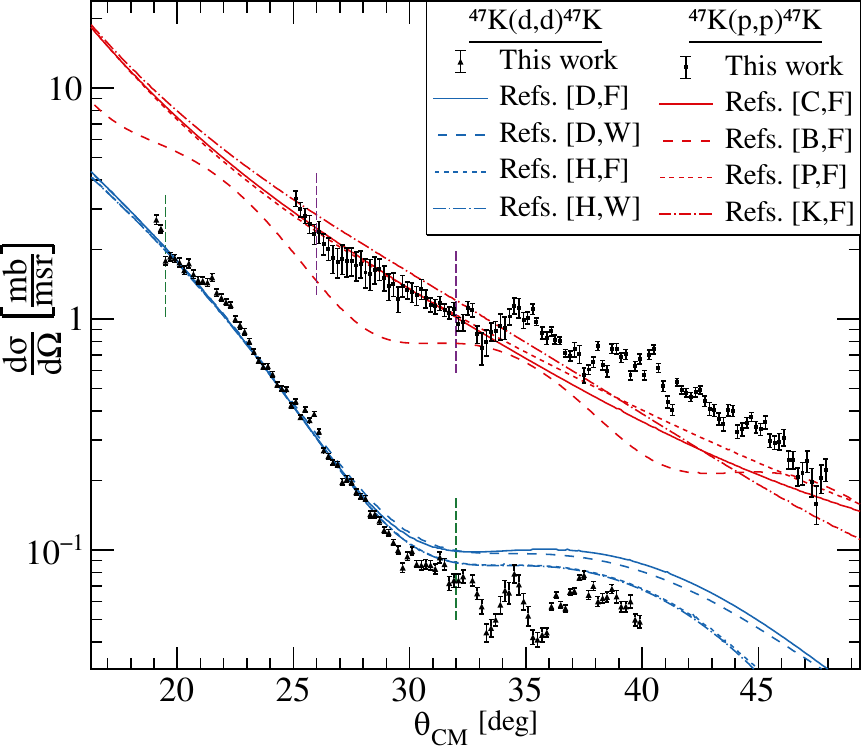}
    \caption{Differential cross sections of experimental elastic scattering data, compared to various optical model calculations. The $^{47}$K(d,d) calculations are shown in blue, and $^{47}$K(p,p) calculations are shown in red. The range of experimental data used in the normalisation is indicated by green and purple dashed lines. Letter codes correspond to the optical model and the computational code. Optical models are from Ref.~\cite{Optical_DCV} (D), Ref.~\cite{Optical_HanShiShen} (H), Ref~\cite{Optical_ChapHill} (C), Ref.~\cite{OpticalModel_BecchettiGreenlees1969} (B), Ref.~\cite{Optical_Perey} (P) and Ref.~\cite{OpticalModel_KoningDelaroche2003} (K). Calculations were performed in FRESCO (F)~\cite{fresco} and DWUCK4 (W)~\cite{dwuck}.}
    \label{fig:elastics}
\end{figure}

Firstly, an analytical approach was taken. The number of observed events in the elastic channels is proportional to the beam intensity, the reaction cross section and the target thickness. The reaction cross section was determined by optical model calculations, performed using the code FRESCO~\cite{fresco} and the potential of Daehnick, Childs and Vrcelj~\cite{Optical_DCV} (Chapel-Hill 89~\cite{Optical_ChapHill}) for deuteron (proton) elastic scattering. These results are shown in Fig~\ref{fig:elastics}. Additional calculations, using alternative optical potentials~\cite{Optical_HSS,Optical_Perey,OpticalModel_BecchettiGreenlees1969,OpticalModel_KoningDelaroche2003} (and in the case of deuteron scattering, an alternative program~\cite{dwuck}) are also shown, demonstrating the close agreement between models over the angular range used for the data fitting. By comparison of experimental elastic scattering and optical model calculations, the normalisation factor $N_d$ ($N_p$) was extracted, which is a product of the number of deuterons (protons) in the target and the integrated flux of the incoming beam. As the incoming beam was monitored with the CATS detector, then the CD$_2$ target thickness and CH$_2$ contamination could be calculated (n.b. Ref.~\cite[pg. 58]{Paxman2024_Thesis}) to be 0.31(2)~mg/cm$^2$ (4\% by number H to D).  

\begin{figure}
    \centering
    \includegraphics[width=\linewidth]{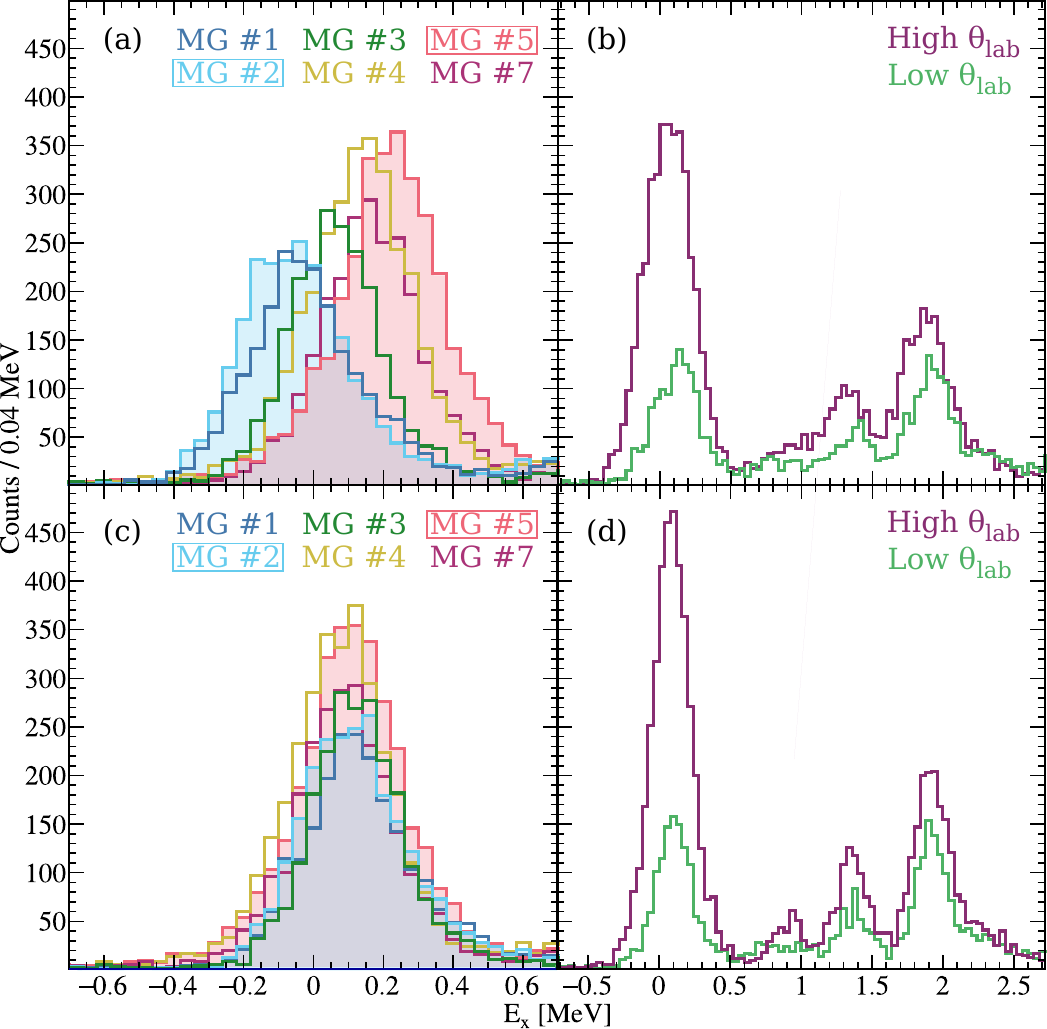}
    \caption{Comparison of the uncorrected (a,b) and corrected (c,d) reconstructed $^{48}$K excitation spectra, when adjusting the beam spot position and target thickness. Spectra in (a,c) are from each of the six upstream detectors. Notably, the corrected spectra show consistent agreement between detectors covering different $\phi_{lab}$ ranges -- see improvement from (a) to (c) -- and different $\theta_{lab}$ ranges -- see improvement from (b) to (d). The lowest-energy peak contains the ground state and 0.143~MeV state, and a such is not centered on 0~MeV.}
    \label{fig:XYCorrection}
\end{figure}

\begin{figure}
    \centering
    \includegraphics[width=\linewidth]{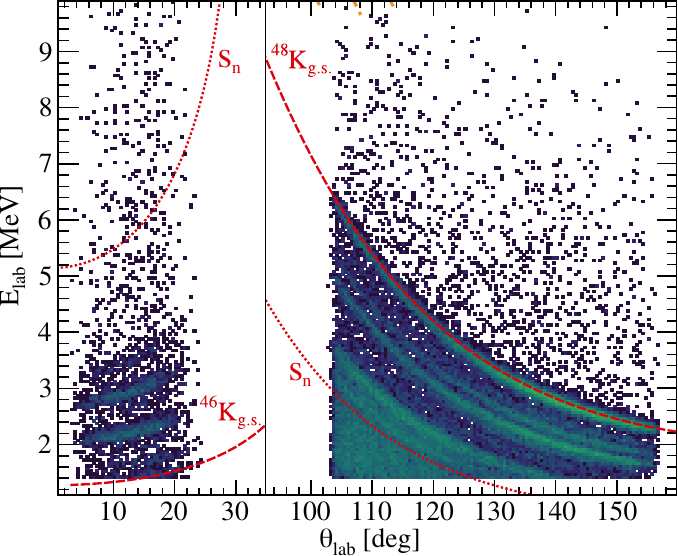}
    \caption{Energy of the detected light ejectiles, E$_{lab}$, versus detected angle in the laboratory frame, for $^{47}$K(d,t)$^{46}$K ($\theta_{lab}=4^\circ-22^\circ$) and $^{47}$K(d,p)$^{48}$K ($\theta_{lab}=104^\circ-156^\circ$). The position of the ground state and the neutron separation energy for each nucleus is marked, and the experimental E$_{lab}$ threshold of 1.4 MeV is visible. Note that the $^{48}$K events shown here only require a timing coincidence from the focal plane of VAMOS++, whereas the $^{46}$K events have additional particle identification requirements to reduce background.}
    \label{fig:ELabThetaLab}
\end{figure}

The target thickness range 0.29 to 0.33~mg/cm$^2$ was taken as the limit of a model-independent, numerical minimization code. The target thickness, target position relative to the detectors, and the average position of the secondary beam on the target were varied during the kinematic reconstruction of the $^{48}$K excitation energy. Coincidence with the 0.143~MeV $\gamma$-ray was enforced, which isolated three states below 2~MeV (0.143, 1.409 and 1.978~MeV) whose precise energies and decays were determined through $\gamma$-ray spectroscopy~\cite{Paxman2025_47Kdp}. These three peaks were fit with Gaussian functions, and the reconstruction of the data was minimized so as to reduce each $\sigma$ whilst maximizing the accuracy of the centroid energies. This resulted in an average beam position of $(-4.16,+0.47)$~mm, which produced the greatest consistency across $\phi_{lab}$ and $\theta_{lab}$ angular ranges. The $^{46}$K data was not used in this fitting due to the increased complexity of the data, but we note a similar consistency with the offset beam position. Using this method, the FWHM of reconstructed excitation energy is 330~keV in $^{48}$K, and 420~keV in $^{46}$K. The impact of the beam spot correction can be seen in Fig.~\ref{fig:XYCorrection}, where the lowest energy peak (containing both the ground state and the unresolved first excited state of $^{48}$K) is properly aligned in each of the upstream MUGAST detectors when the beam offset is applied. Crucially, the reconstructed energies are consistent across ranges of $\theta$ following these corrections, which is critical for the reliable determination of differential cross sections. In addition, the accuracy of the particle-reconstructed excitation energy is in excellent agreement with energies determined by $\gamma$-ray coincidences -- including the states not in the initial fitting algorithm -- and only minor post-hoc corrections of $E_{true}(^{48}$K$)=0.989E_{meas}$ and $E_{true}(^{46}$K$)=E_{meas}+60$~keV are applied, which are small compared to the resolution of the peaks.

The quality of the observed data is clear in Fig.~\ref{fig:ELabThetaLab}, where the large angular coverage and high-statistics measurement of $^{47}$K(d,p) reveals several clear, well-defined kinematic lines relating to the many states populated in $^{48}$K. Also revealed in this figure is the challenge presented by the $^{47}$K(d,t) data, which has fewer total measured events over a smaller angular range. There are several clear kinematic lines relating to excited states, though the kinematic lines of the lowest-energy states are partially obscured by the low energy threshold at $E_{lab}=1.4$~MeV. Notably, there is no discernible population of states at very large energies.
\section{Analysis}

In this paper, we present the results of the first $^{47}$K(d,t)$^{46}$K direct transfer study, followed by a summary of minor states in $^{47}$K(d,p)$^{48}$K not explored in Ref.~\cite{Paxman2025_47Kdp}. In each case, analysis is conducted by first determining the states which are populated through particle-$\gamma$ coincidence spectroscopy. Then, the area of the excitation energy peaks is determined across various sections of the experimental angular range, by fitting multiple Gaussian distributions with fixed excitation energies and widths. The angular sections are $2^{\circ}$ in $\theta_{lab}$ for $^{48}$K, and either $1$ or $2^{\circ}$ in $\theta_{cm}$ for $^{46}$K, depending on the available statistics. The number of observed counts is then corrected by the geometric solid angle as determined by realistic Monte Carlo simulations using GEANT4~\cite{Agostinelli2003_GEANT4} with the nptool~\cite{Matta_2016_nptool} simulation package. The number of efficiency-corrected counts is then scaled by the normalization factor $N_d$, as determined by the simultaneous elastic scattering measurement, to extract the experimental differential cross section, $\frac{d\sigma}{d\Omega}_{\mathrm{exp}}$. Finally, this is compared to the theoretical differential cross-section, $\left(\frac{d\sigma}{d\Omega}\right)_{\mathrm{thry}}$, determined through calculations performed with TWOFNR~\cite{TWOFNR}. The (d,p) calculations employed the Koning-Delaroche global optical potential~\cite{OpticalModel_KoningDelaroche2003} and the Johnson-Tandy adiabatic model (ADWA)~\cite{JohnsonTandyRef}, and (d,t) calculations employed the Daehnick, Childs, and Vrcelj~\cite{Optical_DCV} deuteron optical potential and the Li, Liang and Cai~\cite{li2007global} triton optical potential in the distorted-wave Born approximation (DWBA). The spectroscopic factor ($S$) of each state was then extracted by the $\chi^2$ minimization of the relationship $\left(\frac{d\sigma}{d\Omega}\right)_{\mathrm{exp}} = S\,\left(\frac{d\sigma}{d\Omega}\right)_{\mathrm{thry}}$. In addition to the statistical uncertainty resulting from the $\chi^2$ minimization, a systematic uncertainty of 20\% is applied to these spectroscopic factors, owing to limitations of the reaction models~\cite{JennyLee20percent}.

\subsection{$^{\mathbf{47}}\textrm{K(d,t)}^{\mathbf{46}}\textrm{K}$\label{anal46K}}
\begin{figure*}[t]
    \centering
    \includegraphics[width=0.90\linewidth]{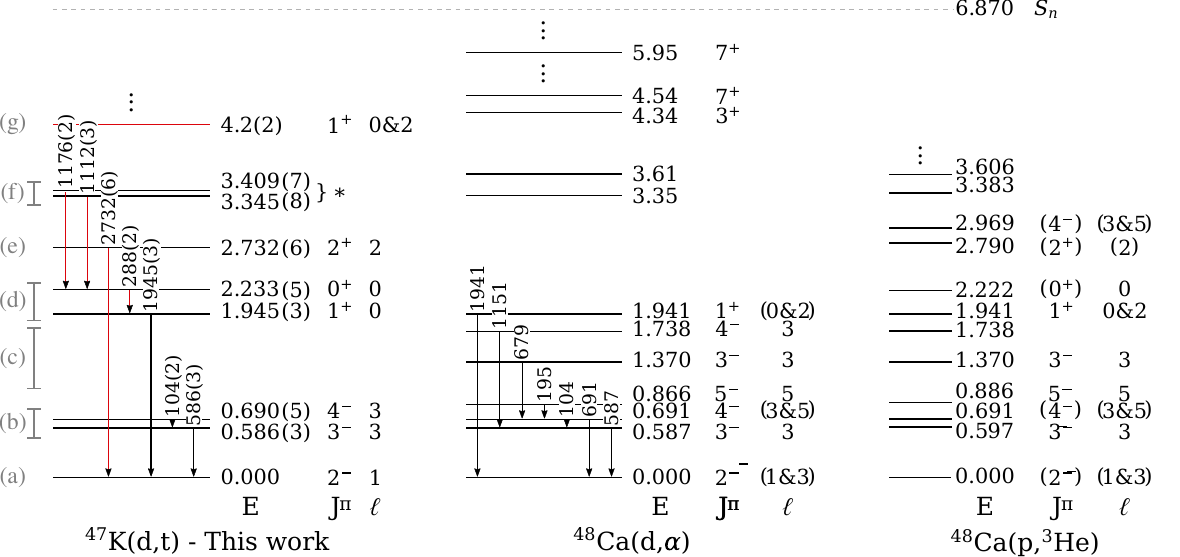}
    \caption{Level scheme of $^{46}$K, as determined through $^{47}$K(d,t) in this work, compared to $^{48}$Ca(d,$\alpha$)~\cite{Paul1971_48Ca-da-46K_NormKine,Daehnick1974_48Ca-da-46K_NormKine} and $^{48}$Ca(p,$^{3}$He)~\cite{Dupont1970_48Ca-p3He-46K_NormKine,Dupont1973_48Ca-p3He-46K_NormKine,Daehnick1973_48Ca-p3He-46K_NormKine} as compiled in Ref.~\cite{NuclearDataSheets_Mass46}. Letter labels relate to the subsection of Section~\ref{anal46K} in which the states(s) are discussed. Four new $\gamma$-ray transitions and a new excited state have been identified in this work, marked in red. The two states marked with an asterisk, labeled (f), are unresolved in particle spectroscopy; the total peak contains $\ell=0$ and $\ell=2$ character, and shell model comparisons would suggest that one state is $0^+$ and the other is $1^+$, but this work is unable to assign these spin-parities to specific states.}
    \label{fig:46K_levelscheme}
\end{figure*}

The analysis of $^{46}$K is built upon an extensive history of direct transfer reactions populating this isotope via $^{48}$Ca(d,$\alpha$)~\cite{Paul1971_48Ca-da-46K_NormKine,Daehnick1974_48Ca-da-46K_NormKine} and $^{48}$Ca(p,$^{3}$He)~\cite{Dupont1970_48Ca-p3He-46K_NormKine,Dupont1973_48Ca-p3He-46K_NormKine,Daehnick1973_48Ca-p3He-46K_NormKine} (see Fig.~\ref{fig:46K_levelscheme}) and expands upon this foundational knowledge as the first measurement of the $^{47}$K(d,t) transfer reaction. Contrary to previous experiments, the states populated through this reaction will be accessed through the \piSOneDFour{} component of their wavefunction, as shown in Fig.~\ref{fig:schematic}(a). As such, neutron removal from the available orbitals could result in states with spin parities of $1^-$ \& $2^-$ from $\nu p_{3/2}^{-1}$, $3^-$ \& $4^-$ from $\nu f_{7/2}^{-1}$, $1^+$ \& $2^+$ from $\nu d_{3/2}^{-1}$ and $0^+$ \& $1^+$ from $\nu s_{1/2}^{-1}$. It cannot be forgotten, however, that the natural ground-state proton configuration of $^{46}$K is \piSTwoDThree{}, and the near-degeneracy of the $\pi s_{1/2}$ and $\pi d_{3/2}$ orbitals leads to additional low-energy states built on this proton configuration. For example, from a simple coupling of the $\pi s_{1/2}^{-1}$$\nu f_{7/2}^{-1}$ orbitals, we would expect to observe one $3^-$ state and one $4^-$ state, but an inspection of the known structure of $^{46}$K~\cite{NuclearDataSheets_Mass46} reveals two such pairs (see Fig.~\ref{fig:46K_levelscheme}). This is explained by the near-degenerate $\pi d_{3/2}$ orbital, which can also couple to $\nu f_{7/2}$ to produce $2^-$, $3^-$, $4^-$ and $5^-$ states at similar energies. It is possible for states of the same spin-parity to mix, so each $3^-$ or $4^-$ state would contain some $\pi s_{1/2}^{-1}$$\nu f_{7/2}^{-1}$ component and some $\pi d_{3/2}^{-1}$$\nu f_{7/2}^{-1}$ component. This would allow the $^{47}$K(d,t) reaction to populate both states of a given $J^{\pi}$ strictly via the $\pi s_{1/2}^{-1}$ component, with the relative strength of their population indicating the degree of proton configuration mixing -- this will be discussed at greater length in Section~\ref{disc_shellmodel}.

\begin{figure*}[t]
    \centering
    \includegraphics[width=\linewidth]{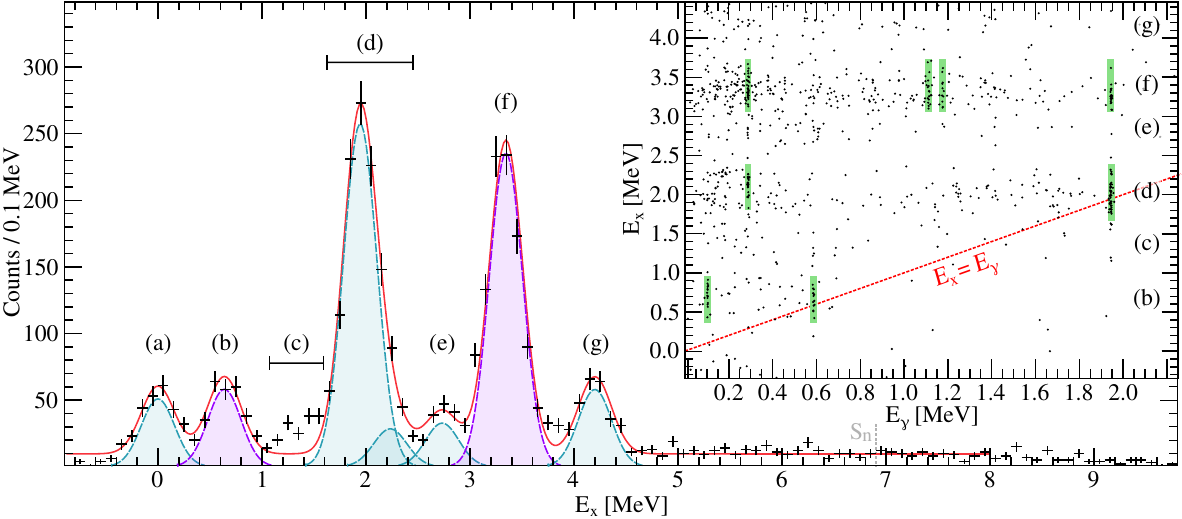}
    \caption{Reconstructed excitation energy spectrum of $^{46}$K data, fitted with Gaussian functions. Individual states are shown as dashed blue lines, doublets as dashed purple lines, and the total fit as a solid red line. Letter labels relate to the subsection of Section~\ref{anal46K} in which the states(s) are discussed. A flat background was fitted in the 5–8~MeV region. The neutron separation energy is indicated by a dashed grey line. (Inset) Excitation energy versus coincident $\gamma$-ray energy, with key particle–$\gamma$ coincidences highlighted in green.}
    \label{fig:46K_Ex-ExEg}
\end{figure*}

The excitation energy spectrum of $^{47}$K(d,t)$^{46}$K is shown in Fig.~\ref{fig:46K_Ex-ExEg}. Notably, there is no observable population of excited states from 5~MeV up to the neutron separation threshold at 6.87~MeV, and beyond to the detection threshold. As such, the region from 5-8~MeV is used to constrain the flat background beneath the populated states. Six peaks are identified in the excitation spectrum -- at 0 MeV, 0.6 MeV, 1.9 MeV, 2.8 MeV, 3.3 MeV, and 4.2 MeV -- but it is clear from $\gamma$-ray coincidences (see inlay of Fig.~\ref{fig:46K_Ex-ExEg}) and from the existing literature~\cite{NuclearDataSheets_Mass46} that several of these peaks are doublets. Here, each peak will be tackled separately.

\paragraph{Ground state}

\begin{figure}
    \centering
    \includegraphics[width=\linewidth]{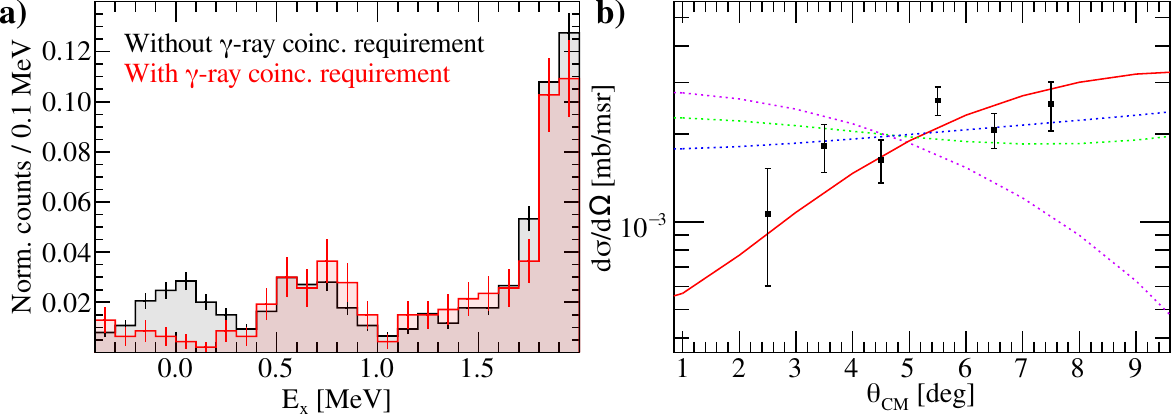}
    \caption{Analysis of $^{46}$K ground state. (a) Normalised counts of $^{46}$K with and without a $\gamma$-ray coincidence requirement. (b) Experimental differential cross section (black) compared to theoretical differential cross sections for $\ell=0$ (purple), $\ell=1$ (red), $\ell=2$ (green) or $\ell=3$ (blue).}
    \label{fig:46K_Ex-0p0MeV}
\end{figure}

The peak centered at 0~MeV is found to contain only the ground state; the requirement of $\gamma$-ray coincidence completely suppresses this peak -- see Fig.~\ref{fig:46K_Ex-0p0MeV}(a) -- and there is no evidence in this work or in the literature for an excited state any lower than 0.5 MeV. This state is analyzed over a smaller angular range than those at higher energies due to the encroaching $E_{lab}$ threshold, as shown in Fig~\ref{fig:ELabThetaLab} -- note that this complication is accounted for by the realistic simulation of the solid angle, wherein the same $E_{lab}$ threshold is applied to the simulated data to ensure reliability. The known $2^-$ spin-parity of the ground state~\cite{Papuga2014} would require that this state be formed by the removal of a $p_{3/2}$ neutron, and indeed, the observed differential cross section state is consistent with $\nu p_{3/2}$ shape, with a spectroscopic factor of $S^{stat}_{sys}=0.34_{0.07}^{0.02}$, as shown in Fig.~\ref{fig:46K_Ex-0p0MeV}(b). The relatively weak stripping of a neutron from $\nu p_{3/2}$ implies a small population of this neutron orbital in the ground state of $^{47}$K, consistent with a slight softness in the $N=28$ shell gap.


\paragraph{0.6~MeV region}

\begin{figure}
    \centering
    \includegraphics[width=\linewidth]{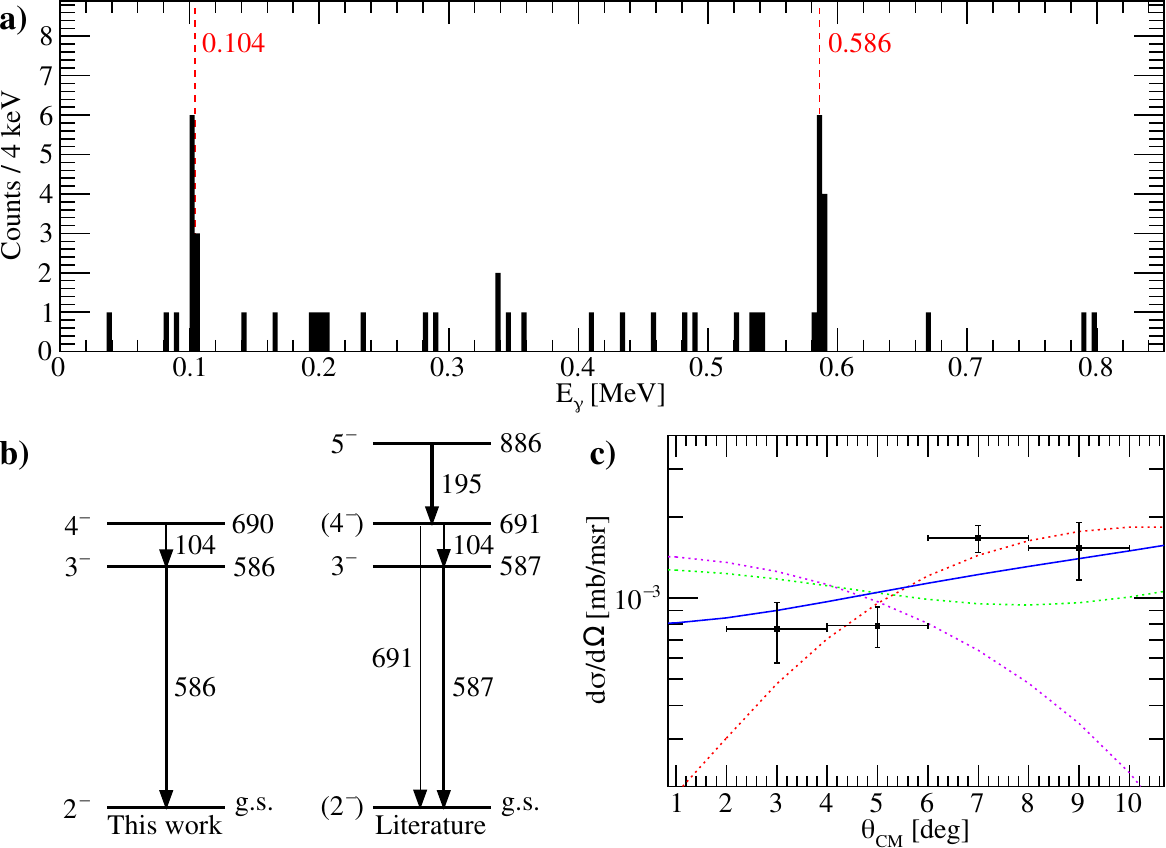}
    \caption{Analysis of the unresolved 0.586 and 0.690~MeV states. (a) Gamma-ray spectrum in coincidence with this peak. (b) Extract of the relevant decay scheme. (c) As Fig.~\ref{fig:46K_Ex-0p0MeV}b.}
    \label{fig:46K_ExEgLS-0p6MeV}
\end{figure}

The peak centered at 0.6~MeV could contain two unresolved states, identified in the literature as 0.587~MeV, $3^-$ and 0.690~MeV, $4^-$. The higher-energy of these states is directly populated by the $^{47}$K(d,t) reaction, as determined by the present observation in Fig.~\ref{fig:46K_ExEgLS-0p6MeV} of two $\gamma$-rays in coincidence with this peak, at 0.586(3)~MeV and 0.104(2)~MeV, in agreement with the literature decays of these states. Critically, the strength of the direct population of the 0.586~MeV state is unclear, due to the large uncertainties in determining the number of 0.586~MeV $\gamma$-rays that originate from direct population, or from feeding of the doublet partner. We do not observe the 0.691~MeV decay reported in Ref.~\cite{Daehnick1974_48Ca-da-46K_NormKine} with a branching ratio of 28\%, which is not unexpected, given the number of 0.104~MeV counts observed, and the comparative efficiencies at the two energies. Finally, we do not expect to observe the known 0.886~MeV state, as this has a spin-parity of $5^-$, and hence is inaccessible by this reaction; indeed, the present non-observation of the known 0.195~MeV decay supports this conclusion.

The $3^-$ \& $4^-$ spin-parities of these states are only accessible by $\nu f_{7/2}$ removal, and indeed, by fitting these unresolved states with a single peak, a differential cross section is observed that is consistent with $\nu f_{7/2}$ removal (see Fig.~\ref{fig:46K_ExEgLS-0p6MeV}c). A similarity with $\nu p_{3/2}$ removal is discounted by spin-parity arguments. As the relative population of the two states cannot be determined by $\gamma$-decay arguments, we here report a combined spectroscopic factor of $S^{stat}_{sys}=7.0^{0.5}_{1.4}$. Hence, this doublet appears to exhaust almost all of the possible $\nu f_{7/2}$ strength.


\paragraph{1.3~MeV to 1.7~MeV region}

A second pair of states is known in the literature in this region -- 1.370~MeV, $3^-$ and 1.738~MeV, $4^-$ -- and present a challenge in this work. These states are accessible in this reaction via $\nu f_{7/2}$ removal, and we would expect these states to be populated via their mixed-proton \piSOneDFour{} component, as discussed in the introduction to Section~\ref{anal46K}. Despite this, evidence for direct population of these states is sparse. 

There is an excess of counts at approximately 1.3~MeV (see Fig~\ref{fig:46K_Ex-ExEg}), but the $\gamma$-ray coincidences do not support the direct population of this state. Looking to the inlay of Fig~\ref{fig:46K_Ex-ExEg}, we would expect to see the known $\gamma$-ray cascade 0.679$\rightarrow$0.104$\rightarrow$0.587~MeV in coincidence with the 1.3~MeV excitation region; while one could argue for the presence of 0.587~MeV, there is no clear cascade. If the excess of counts were to arise from population of the 1.370~MeV state, we would expect to observe approximately 15 counts at 0.104~MeV and 9 counts at 0.587 and 0.679~MeV, respectively, which the results are far short of. Evidence for the 1.738~MeV state is even weaker, with no clear $\gamma$-ray coincidences in this region.

To assess this further, we estimate the maximum possible strength of this excess of counts by comparison to the neighboring 0.6~MeV peak -- assuming $\nu f_{7/2}$ character, and correcting for the differing solid angle coverage of the two energetic regions. Through this method, we find an upper limit on the spectroscopic factor of $S<3$, however it should be noted that the 0.6~MeV doublet exhausts 88(24)\% of the $\nu f_{7/2}$ strength.


\paragraph{1.9~MeV region} 

\begin{figure}
    \centering
    \includegraphics[width=\linewidth]{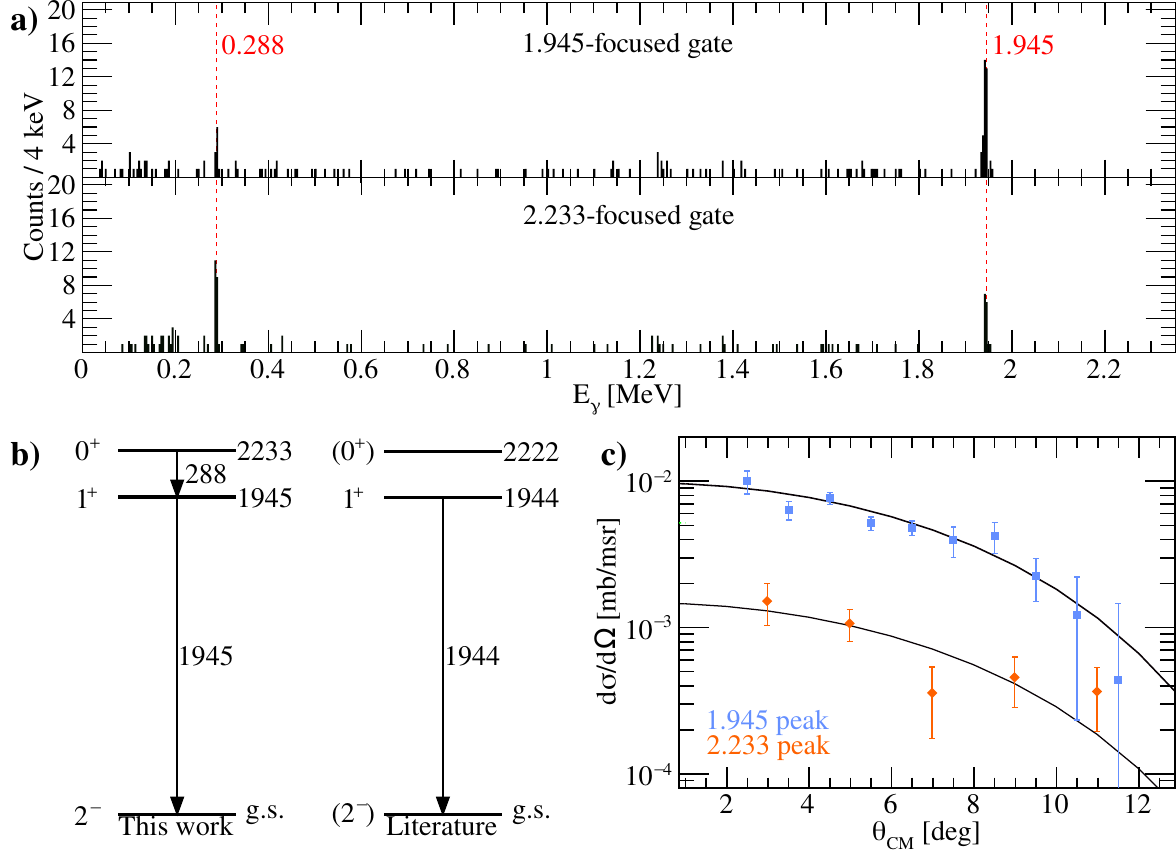}
    \caption{Analysis of the 1.945 and 2.233~MeV states. (a) Gamma-ray spectrum in coincidence with the low- and high-energy sides of this peak. (b) Extract of the relevant decay scheme. (c) Experimental differential cross sections of 1.945~MeV (blue) and 2.233~MeV (orange) compared to theoretical differential cross sections for $\ell=0$ (black).}
    \label{fig:46K_ExEgLS-1p9MeV-2p2MeV}
\end{figure}

The peak centered at 1.9~MeV contains two unresolved states; a strong state at 1.945(3)~MeV, and a weaker state at approximately 2.2~MeV. In the literature, both of these states were known, with a decay from 1.945~MeV, $1^+$ to the ground state reported, and the higher-energy state reported as 2.222~MeV, ($0^+$) from pure particle spectroscopy. Through present particle-$\gamma$ coincidence, shown in Fig.~\ref{fig:46K_ExEgLS-1p9MeV-2p2MeV}, a 0.288(2)~MeV $\gamma$-decay between the states has been observed for the first time, revising the energy of the higher-energy state to 2.233(5)~MeV. As the particle states are not well-resolved, the $\gamma$-decay path is verified by comparing the spectra when gating on the low-energy side of the peak, which suppresses the 0.288~MeV decay, and the high-energy side of the peak, which enhances this decay. This effect can also be seen in the inlay of Fig~\ref{fig:46K_Ex-ExEg}.

In this case, the separation of the two states is sufficient for separate differential cross-section extractions, which are shown in Fig.~\ref{fig:46K_ExEgLS-1p9MeV-2p2MeV}(c). The strong 1.945~MeV state has a clear $\nu s_{1/2}$ character, consistent with the established $1^+$ spin-parity -- notably, this has been assigned as a mixed $\ell=0+2$ state in previous multi-nucleon removal experiments~\cite{Daehnick1974_48Ca-da-46K_NormKine}, which is not clearly apparent in this work. The weaker 2.233~MeV state, despite poorer statistics, is also consistent with $\nu s_{1/2}$ removal, supporting the $(0^+)$ literature assignment. Here, we report spectroscopic factors of $S^{stat}_{sys}=0.56^{0.02}_{0.11}$ and $S^{stat}_{sys}=0.10^{0.01}_{0.02}$ for the 1.945(3)~MeV and 2.233(5)~MeV states, respectively.


\paragraph{2.732~MeV state}

\begin{figure}
    \centering
    \includegraphics[width=\linewidth]{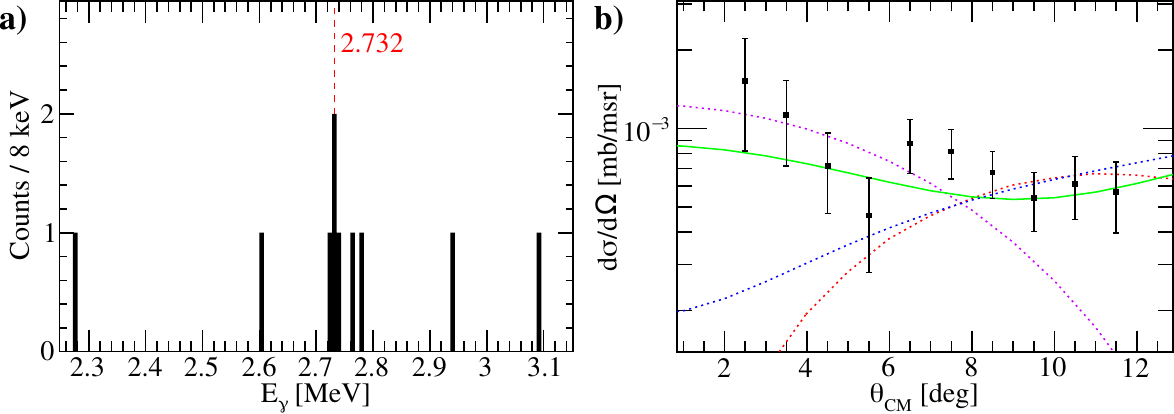}
    \caption{Analysis of the 2.732~MeV state. (a) Gamma-ray spectrum in coincidence with this peak. (b) Experimental differential cross section (black) compared to theoretical differential cross sections for $\ell=0$ (purple), $\ell=1$ (red), $\ell=2$ (green) or $\ell=3$ (blue).}
    \label{fig:46K_ExEgLS-2p7MeV}
\end{figure}

A single state is observed at 2.73~MeV, and particle-$\gamma$ coincidence gating, shown in Fig.~\ref{fig:46K_ExEgLS-2p7MeV}(a), reveals a tentative decay directly to the ground state, establishing the energy of this excited state as 2.732(6)~MeV. 
The differential cross section of this state, shown in Fig.~\ref{fig:46K_ExEgLS-2p7MeV}(b), is only consistent with $\nu d_{3/2}$ transfer, limiting the possible spin-parities of the state to $1^+$ or $2^+$. While the energy of this state does not match exactly with the 2.79~MeV $2^+$ state reported in the literature, that energy was determined solely through particle spectroscopy, and a variation of 50 keV is not unrealistic. Additionally, comparison to shell model calculations (presented in Section~\ref{disc_shellmodel}) support the conclusion that this state is of $2^+$ character, by energy and $\ell$-transfer matching arguments. A spectroscopic factor of $S^{stat}_{sys}=1.6^{0.2}_{0.3}$ is extracted for this state -- note the large statistical error, which results from the relative flatness of the $\nu d_{3/2}$ differential cross section over this angular range.


\paragraph{3.3~MeV region} 

\begin{figure}
    \centering
    \includegraphics[width=\linewidth]{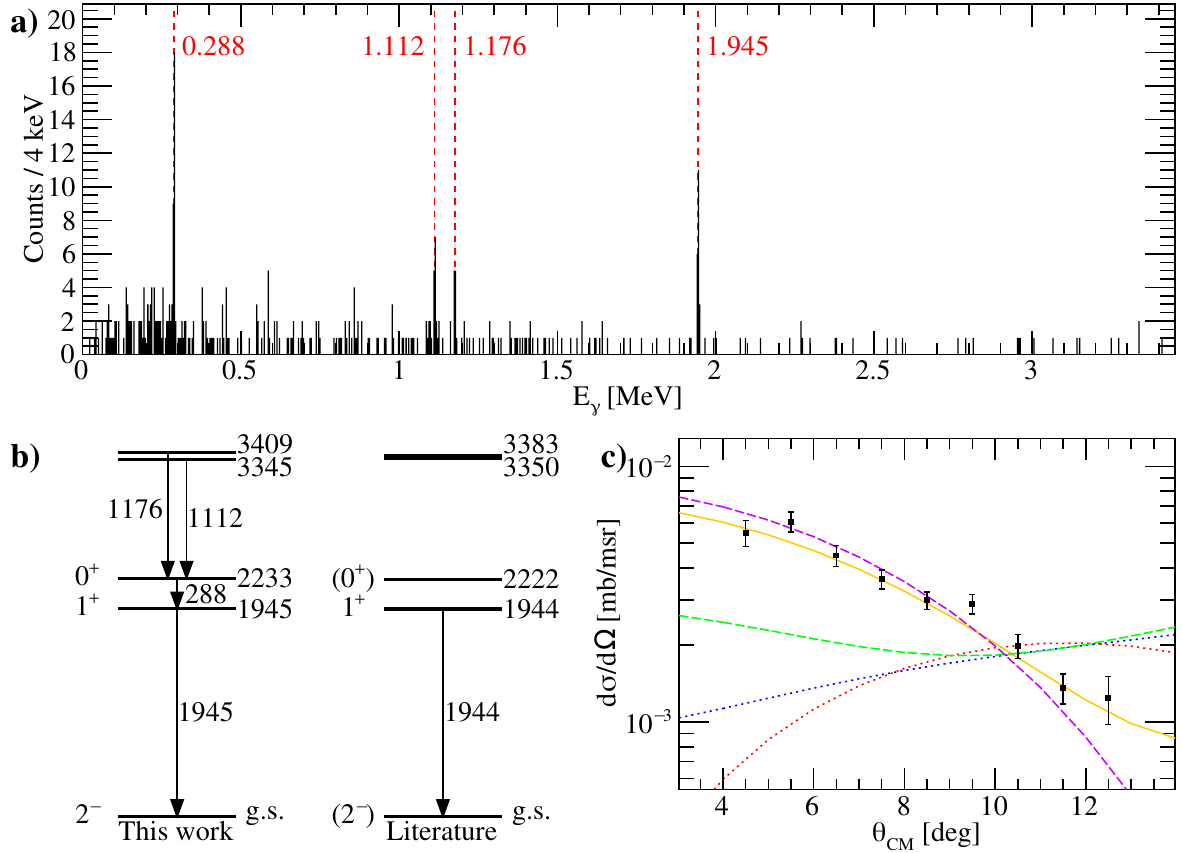}
    \caption{Analysis of the 3.344 and 3.410~MeV unresolved states. As Fig.~\ref{fig:46K_ExEgLS-0p6MeV}, but with the addition in (c) of the $\ell=0+2$ mixed configuration differential cross section (orange). }
    \label{fig:46K_ExEgLS-3p3MeV}
\end{figure}

In the study of $^{48}$Ca(p,$^{3}$He)$^{46}$K by Daehnick \textit{et al.}~(1973)~\cite{Daehnick1973_48Ca-p3He-46K_NormKine}, an unresolved doublet is reported at 3.383 MeV. Here, through particle-$\gamma$ coincidence spectroscopy, we report two new $\gamma$-ray decays observed in coincidence with this strongly-populated region, at 1.112(3)~MeV and 1.176(2)~MeV. Additionally, the observation of the 1.945~MeV and 0.288~MeV $\gamma$-rays indicate that both of the new transitions are to the 2.233(5)~MeV state, affixing the energies of the two unresolved states as 3.345(8)~MeV and 3.409(7)~MeV. This interpretation is supported by the unpublished results~\footnote{E. Clément (personal communication)} of a high-statistic $\gamma$-ray study of the $^{46}$Ar beta decay, which found the 1.112~MeV and 1.176~MeV $\gamma$-rays to both be in coincidence with the 0.288-1.945~MeV chain. 

The differential cross section of this doublet is shown in Fig~\ref{fig:46K_ExEgLS-3p3MeV}(c), and shows a slight mixed character. While the $\nu s_{1/2}$ strength is obvious, the relative flatness of the $\nu d_{3/2}$ shape presents difficulty when assessing mixing, leading to large statistical errors. It is expected, through comparison to shell model calculations which will be discussed further in Section~\ref{disc_shellmodel}, that this doublet would contain a $0^+$ state and a $1^+$ state, but we are unable to make any determination regarding spin-parity from this work. Given the combination of the unknown spin-parities and $\nu(s_{1/2}d_{3/2})$ mixing, we are unable to determine how this strength is distributed between the two states. As such, we report here a combined spectroscopic factor of $S^{stat}_{sys}$($\nu s_{1/2}$)=$0.79^{0.08}_{0.16}$ and $S^{stat}_{sys}$($\nu d_{3/2}$)=$2.3^{0.7}_{0.5}$ for the 3.345(8) and 3.409(7)~MeV states. 


\paragraph {4.2~MeV state}

A new, well-resolved state is observed at 4.2~MeV. Unfortunately, there are no clear particle-$\gamma$ coincidences between the 4.2~MeV state and the known $\gamma$-decay structure of $^{46}$K; see, for example, the inset of Fig.~\ref{fig:46K_Ex-ExEg}. In considering a decay directly to the ground state, based on the 240(20) observed particle events and the AGATA detection efficiency $\epsilon_{\mathrm{AGATA}}(4.2~\mathrm{MeV})=3.3\%$, we would expect approximately eight particle-$\gamma$ coincidence events -- in fact, we observe no $\gamma$-ray coincidences at this energy, but we note that the absence of an observed ground-state decay in the present data does not necessarily exclude its occurrence. 

\begin{figure}
     \centering
     \includegraphics[width=0.8\linewidth]{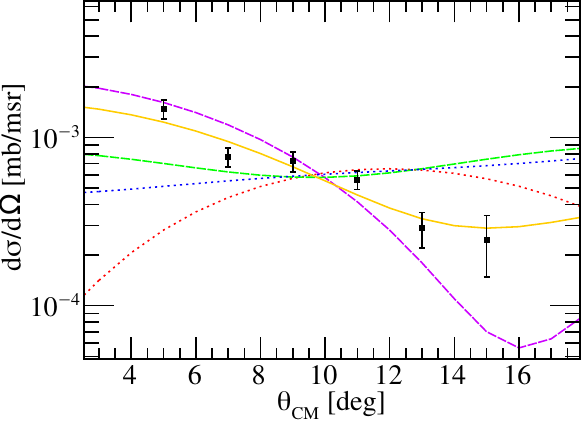}
     \caption{As Fig.~\ref{fig:46K_ExEgLS-3p3MeV}(c), for the novel 4.2~MeV state, showing a clear mixed $\ell=0+2$ configuration (orange).}
     \label{fig:Results_46K_4p2}
\end{figure}

Contrary to the limited $\gamma$-decay spectroscopy of this state, the particle spectroscopy is particularly revealing, as shown in Fig.~\ref{fig:Results_46K_4p2}. The experimental differential cross section of the 4.2 MeV state is of good quality, with a broad angular range and small statistical errors, but it is not well-described by any of the four pure-$\ell$ theoretical curves. Instead, this state exhibits a convincing mixed $\nu(sd)$ character, with $S^{stat}_{sys}$($\nu s_{1/2}$)=$0.19^{0.06}_{0.04}$ and $S^{stat}_{sys}$($\nu d_{3/2}$)=$1.5^{0.5}_{0.3}$. As only $1^+$ states can carry both $\nu s_{1/2}$ and $\nu d_{3/2}$ strength in this reaction, this is strong evidence for a $1^+$ spin-parity assignment of this novel state.

\subsection{$^{\mathbf{47}}\textrm{K(d,p)}^{\mathbf{48}}\textrm{K}$\label{anal48K}}
\begin{figure*}
    \centering
    \includegraphics[width=\linewidth]{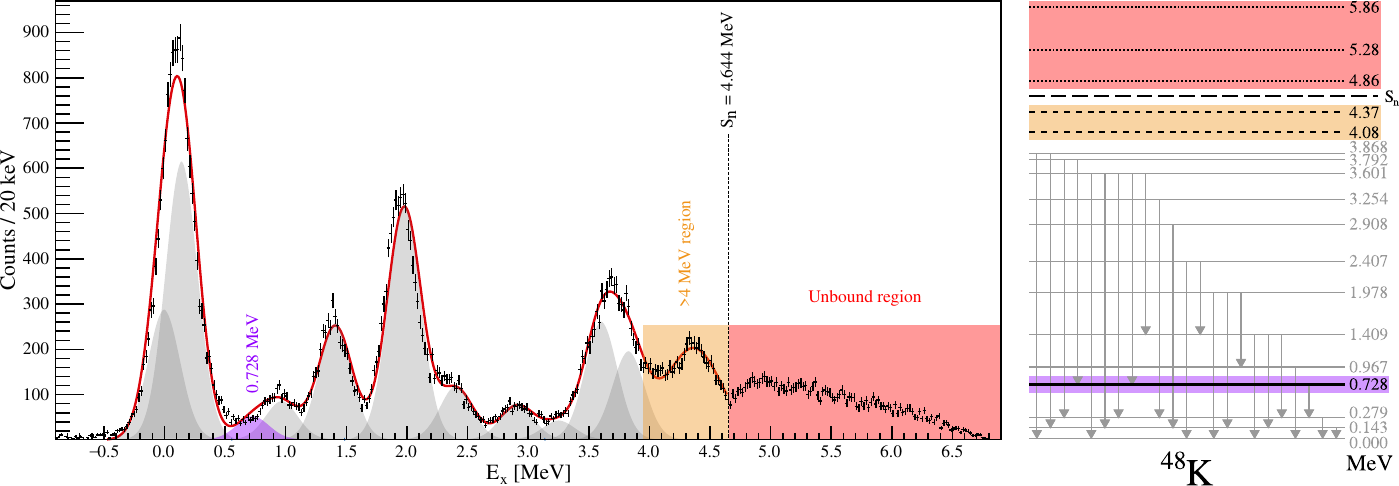}
    \caption{Excitation spectrum (left) and associated level scheme (right) of $^{48}$K, as determined through $^{47}$K(d,p). The three regions focused on in this paper -- the weak 0.728 MeV state, the high energy bound region, and the unbound region -- are highlighted, with the states analyzed in Ref.~\cite{Paxman2025_47Kdp} shown in grey.}
    \label{fig:48K_SeparateExEg}
\end{figure*}

\begin{figure}
    \centering
    \includegraphics[width=\linewidth]{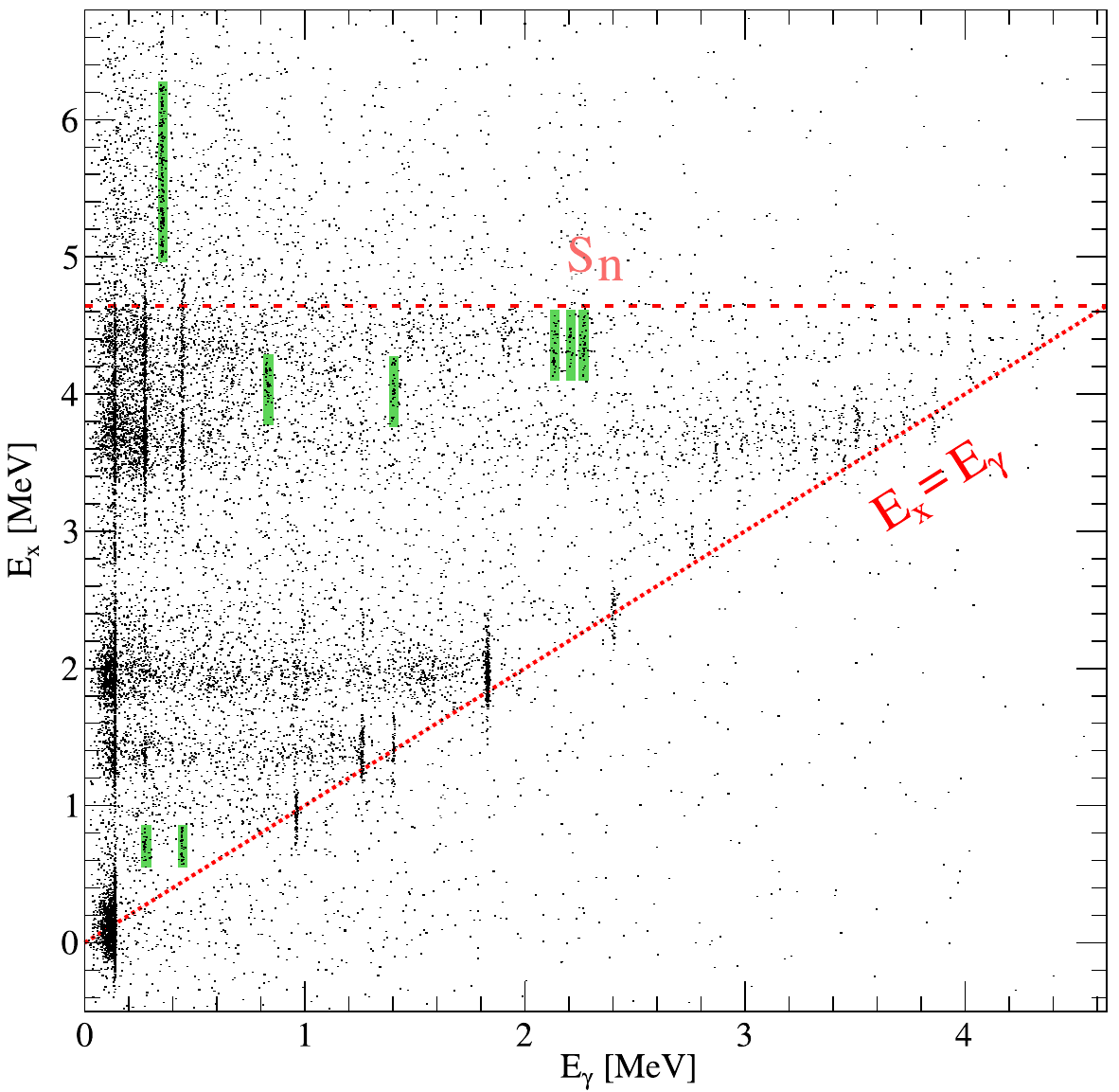}
    \caption{Observed \ce{^{47}K}(d,p)\ce{^{48}K} excitation and $\gamma$-ray transition coincidences, with some features discussed in-text indicated in green. The neutron separation energy (red, dashed) and the line of excitation-decay unity (red, solid) are indicated.}
    \label{fig:48K_2DParticleGamma}
\end{figure}

Turning now to the $^{47}$K(d,p) reaction, we here expand on the analysis presented in Ref.~\cite{Paxman2025_47Kdp}, with details of the very weak states, complex states, and states beyond the neutron separation threshold. These states are highlighted in Fig.~\ref{fig:48K_SeparateExEg}. In order to begin discussion of these difficult states, we turn first to the two-dimensional comparison of the energy of the $^{48}$K state and the de-exciting $\gamma$-ray transitions out of that state, shown in Fig.~\ref{fig:48K_2DParticleGamma}. This figure highlights three key points of interest in the discussion of weak or unusual states in $^{48}$K; that is, the 0.728~MeV state, the 4~MeV region, and the unbound region.

\paragraph{0.728~MeV state}

In Fig.~\ref{fig:48K_2DParticleGamma}, two weak coincidences with the 0.279(1)~MeV and 0.449(2)~MeV $\gamma$-rays are observed at $E_{x}\sim0.7$, which is consistent with the known $\gamma$-ray cascade of the 0.728(3)~MeV, $3^-$ state, evidencing the direct population of this state. This state is expected to be weakly populated by transfer into $\nu f_{7/2}$ orbital from the small amount of $\pi s_{1/2}^{-1} \nu(f_{7/2}^{-2}p_{3/2}^2)$ 2p-2h cross-shell excitation in the ground state of $^{47}$K; indeed, the efficiency-corrected number of observed $\gamma$-rays is very small, with 260(50) counts at 0.279~MeV and 330(60) counts at 0.449~MeV. Notably, there is no evidence of direct population of the 0.279~MeV, $2^-$ state, which could be populated via the $\nu p_{3/2}$ orbital. In Ref.~\cite{Paxman2025_47Kdp}, we reported upper limits on the spectroscopic factors of the 0.279~MeV and 0.728~MeV states, based on the maximum possible from particle spectroscopy. Here, we clarify those results based on $\gamma$-spectroscopy presented in Fig.~\ref{fig:48K_2DParticleGamma}, with no discernible direct population of the 0.279~MeV state, and a very small direct population of the 0.728~MeV state. The spectroscopic factor of the 0.728~MeV state, assuming it is populated via $\nu f_{7/2}$, is found to be 0.05(2).

\paragraph{4~MeV region}

In Ref.~\cite{Paxman2025_47Kdp}, the strength in the region from 4.0-4.6~MeV was treated as one broad region, owing to several inextricable states. Here, we present the particle-$\gamma$ spectroscopy of this region in more detail. Turning first to Fig.~\ref{fig:48K_2DParticleGamma}, we observe five clear $\gamma$-rays in this region which do not have any coincidences with other excitation energies; 0.836(3)~MeV and 1.411(4)~MeV centered at 
$E_x=4.08(6)$~MeV, and 2.140(4)~MeV, 2.210(6)~MeV and 2.274(3)~MeV centered at 
$E_x=4.37(3)$~MeV. In addition, there are strong coincidences with the 0.143, 0.279 and 0.449~MeV $\gamma$-rays, which originate from the first three excited states in $^{48}$K, but there is a clear lack of coincidences with $\gamma$-rays originating from other intermediate states. As such, it appears that these states do not decay significantly via the observed states in the region 0.8-3.8~MeV. This presents a challenge, as the sum of the observed $\gamma$-ray decays is not sufficient without an additional intermediate state; e.g. a cascade of $0.836 + 1.411 + 0.279 + 0.449$~MeV could not originate from a state around $4.1$~MeV. Unfortunately, no candidate intermediary states could be identified. 

In the case of the 2.140(4)~MeV, 2.210(6)~MeV and 2.274(3)~MeV decays, any combination of two of the transitions could be a cascade from a $\sim$4.4 MeV state to the ground state, based on energy summing arguments. Turning to the comparative intensities of the three $\gamma$-ray transitions, it is possible that the 2.274~MeV decay, with 740(120) efficiency-corrected counts, could be fed by both the 2.140~MeV, 590(100) counts, and the 2.210~MeV, 410(90) counts, which would establish two excited states at 4.414(7)~MeV and 4.484(9)~MeV, respectively. However, this does not account for the clear coincidence with the low-lying 0.143, 0.279 and 0.449~MeV $\gamma$-rays.

There are several possible reasons for the apparent incomplete $\gamma$-cascades. It could be that the observed decays are to some state(s) with non-negligible lifetime(s), wherein the isomer decay would not be detected by AGATA, or alternatively; the intermediate state could itself have very fractured, low-intensity decay pathways, so each transition is less likely to be observable. In this context, we note the presence of a $5^+$ isomer at 2.177~MeV with lifetime on the order of nanoseconds, observed in previous studies of $^{48}$K~\cite{Krolas_2011}, which would be at the right approximate energies to be a companion to one of the 4.4~MeV decays, but we also caution that a strong decay to a positive-parity state from the negative-parity states populated in this transfer reaction would be unusual. Ultimately, the complexity of this data prevents us from drawing conclusions about the structure in this high-energy region without more detailed $\gamma$-ray spectroscopy of $^{48}$K.

\paragraph{Unbound states}

\begin{figure}
    \centering
    \includegraphics[width=\linewidth]{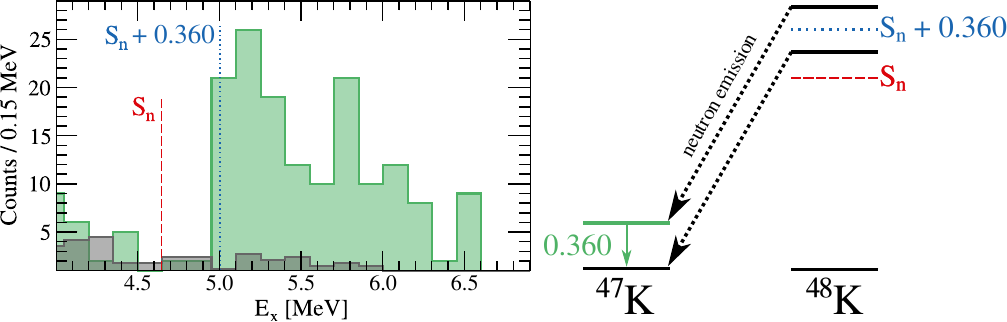}
    \caption{Excitation spectrum observed in coincidence with the 0.360~MeV $\gamma$-ray from the first excited state of $^{47}$K (green), and a selected background region (grey). Note the significant strength at excitation energies greater than the neutron separation energy, $S_n$, summed with the $^{47}$K excited state energy, consistent with narrow unbound states which decay by neutron emission and subsequent $\gamma$-ray de-excitation.}
    \label{fig:Results_48K_GammaGateUnboundState}
\end{figure}

The events above the $^{48}$K neutron separation threshold, $\mathrm{S_n} = 4.644$~MeV, are largely attributed to the three-body deuteron break up channel $^{47}$K\,+\,d~$\rightarrow$~$^{47}$K\,+\,p\,+\,n, characterized using realistic Monte Carlo simulations in GEANT4~\cite{Agostinelli2003_GEANT4} and the simulation package of \textit{nptool}~\cite{Matta_2016_nptool}, however there are events that are not accounted for by the bound states or by the breakup of deuterons. In addition, particle-$\gamma$ coincidences (Fig.~\ref{fig:48K_2DParticleGamma}) reveal a clear feature in the unbound region in coincidence with $E_{\gamma}=0.36(1)$~MeV. This $\gamma$-ray is not observed in coincidence with any bound states of $^{48}$K, and is in fact associated with the decay of the first excited state in $^{47}$K. Hence, this feature is clear evidence for the population of unbound states in $^{48}$K, which decay by neutron emission to $^{47}$K*, and then $\gamma$-decay to $^{47}$K$_{\textrm{g.s.}}$. This is exemplified in Fig.~\ref{fig:Results_48K_GammaGateUnboundState}, which shows clear unbound structures. While $\gamma$-ray coincidences in the unbound region only begin above $\sim$5~MeV, the particle-only spectrum indicates that there is also an excess of counts between $S_{n}=4.644$~MeV and 5~MeV. This would be consistent with an unbound state which cannot decay to the first excited state of $^{47}$K, as the state is unbound by less than 0.36~MeV.

To isolate the unbound states, the following procedure was performed (n.b. Ref.~\cite[pg.87]{Paxman2024_Thesis}). Firstly, the unbound states occupy a small triangular region in $E_{lab}$-$\theta_{lab}$ spectrum shown in Fig.~\ref{fig:ELabThetaLab}, flanked on two sides by the $E_{lab}$ threshold and the physical edge of the detectors. In an effort to remove possible confounding edge effects, a narrow rectangular two-dimensional $E_{lab}$-$\theta_{lab}$ gate was applied to the experimental and simulated data, wherein the intersections of the $E_x=6.0$~MeV kinematic line with the $E_{lab}$ and $\theta_{lab}$ were taken as corners, and projected parallel to the line of increasing $E_{x}$. This gate recovers the expected increasing slope shape of the deuteron breakup background without the shaping effects of the physical detector coverage.

\begin{figure}
    \centering
    \includegraphics[width=\linewidth]{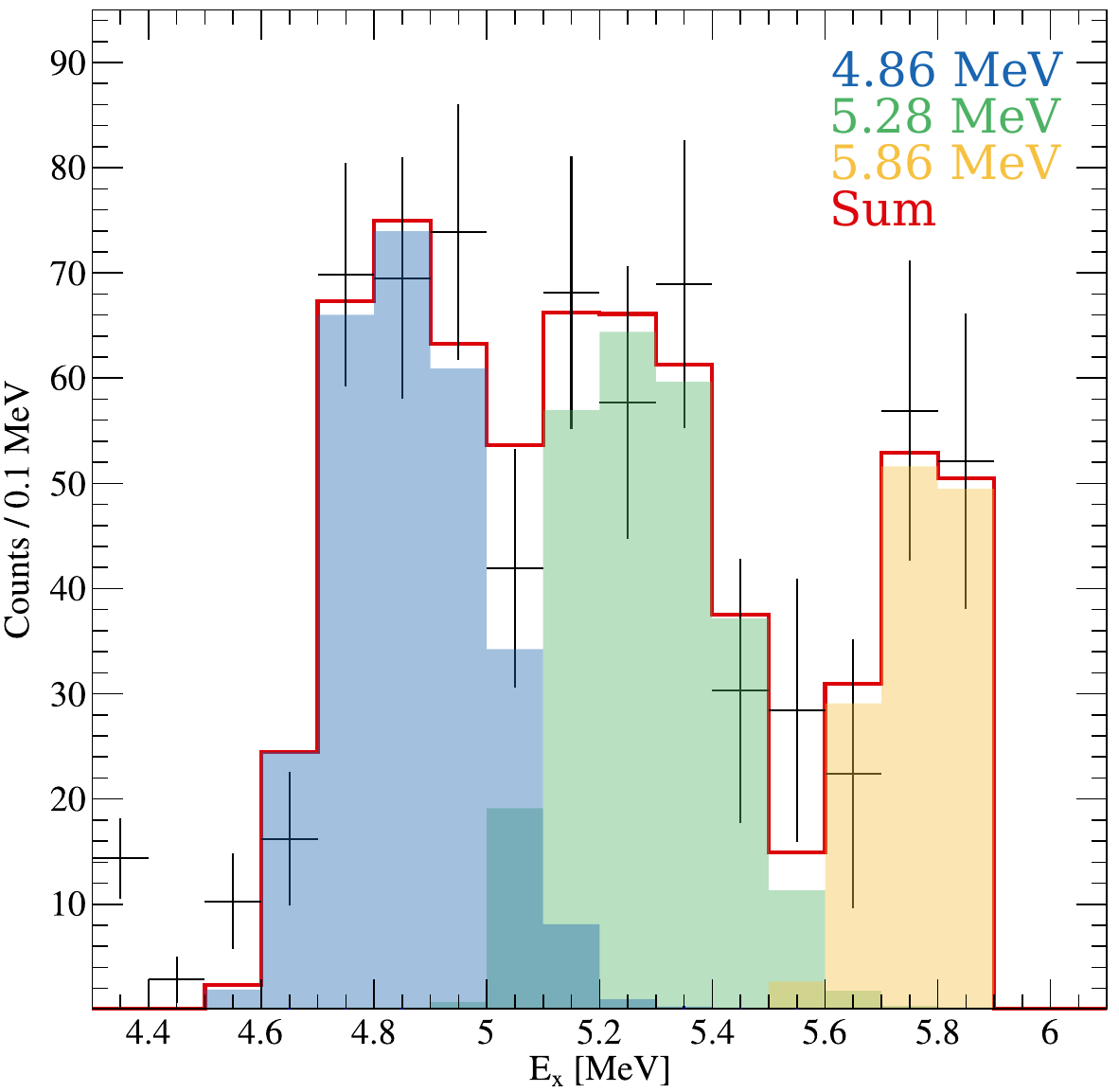}
    \caption{Unbound states in $^{48}$K, fit with three simulated peaks. Note the high-energy cut-off at 6.0~MeV, owing to the two-dimensional $E_{lab}$$-\theta_{lab}$ gate described in-text.}
    \label{fig:48K_UnbFits}
\end{figure}

Using this gate, the known bound states were fit and subtracted, as the tail of the highest energy bound states could alter the shape of the low-energy unbound region. Then, the simulated deuteron breakup background was maximally subtracted. The remaining events, shown in Fig.~\ref{fig:48K_UnbFits}, are hence attributed to unbound states. Further simulations were performed for transfer into unbound states between 4.63~MeV and 6.20~MeV, in steps of 0.02~MeV. Through $\chi^2$ minimization, the best fit to the data was found using three simulated peaks at 4.86, 5.28 and 5.86~MeV, indicating three unbound states at 4.9(2), 5.3(2), and 5.9(2)~MeV.

Unfortunately, due to the small angular range over which these states were observed, no differential cross section could be extracted. Instead, the area of each peak was evaluated (with an additional error of 10\% applied owing to uncertainty in peak centroid energies), adjusted for solid angle efficiency, and compared to optical model calculations, performed with TWOFNR~\cite{TWOFNR} using the Koning-Delaroche global optical potential~\cite{OpticalModel_KoningDelaroche2003} and
Johnson-Tandy adiabatic model~\cite{JohnsonTandyRef}, in accordance with the other states observed in this work and Ref.~\cite{Paxman2025_47Kdp}. Note that, due to difficulty in optical model calculations of unbound states, we here adopt the weakly-bound neutron approximation~\cite{Sen1974_UnbndDWBA_CompareVincFortToWeakBound}. By comparison to shell-model calculations, these states are most probably $f_{5/2}$ in character, though no discrimination between $2^-$ or $3^-$ spin-parity is possible. If the 4.9(2), 5.3(2) and 5.9(2)~MeV states were $2^-$ in character, they would carry spectroscopic factors of 0.07(3), 0.07(3) and 0.06(2) respectively. Alternatively, if they were $3^-$ in character, they would have spectroscopic factors of 0.05(2), 0.05(2) and 0.04(2), respectively. An uncertainty of 40\% has been placed on these values, owing to error introduced by peak fitting (10\%), phase space subtraction (5\%) and the weakly-bound neutron approximation (5\%), in addition to the usual 20\% error associated with the choice of optical model.

\section{Discussion}

\subsection{Shell model calculations of $^{47}$K(d,t)$^{46}$K\label{disc_shellmodel}}

In order to interpret these experimental results, shell model calculations for the $^{47}$K(d,t) reaction were performed by B. A. Brown~\footnote{B.\,A.~Brown (personal communication)}, using an adapted ZBM2* Hamiltonian~\cite{Brown2022_ZBM2*}. This calculation was performed in a model space consisting of $\pi(2s_{1/2},1d_{3/2},1f_{7/2},2p_{3/2})$ and $\nu(2s_{1/2},1d_{3/2},1f_{7/2},2p_{3/2})$, which covers the appropriate neutron orbitals, and -- crucially -- can account for excitations across the $N=28$ shell gap, originating from the diffuse neutron occupancy distribution near the Fermi surface. The exclusion of the deep $\nu(1d_{5/2})$ orbital is not concerning, particularly given the clear absence of excited states at very high energies in Fig.~\ref{fig:46K_Ex-ExEg}, and the relatively good matching in terms of the number and energy of states predicted, as discussed below.

 \begin{figure}
     \centering
     \includegraphics[width=\linewidth]{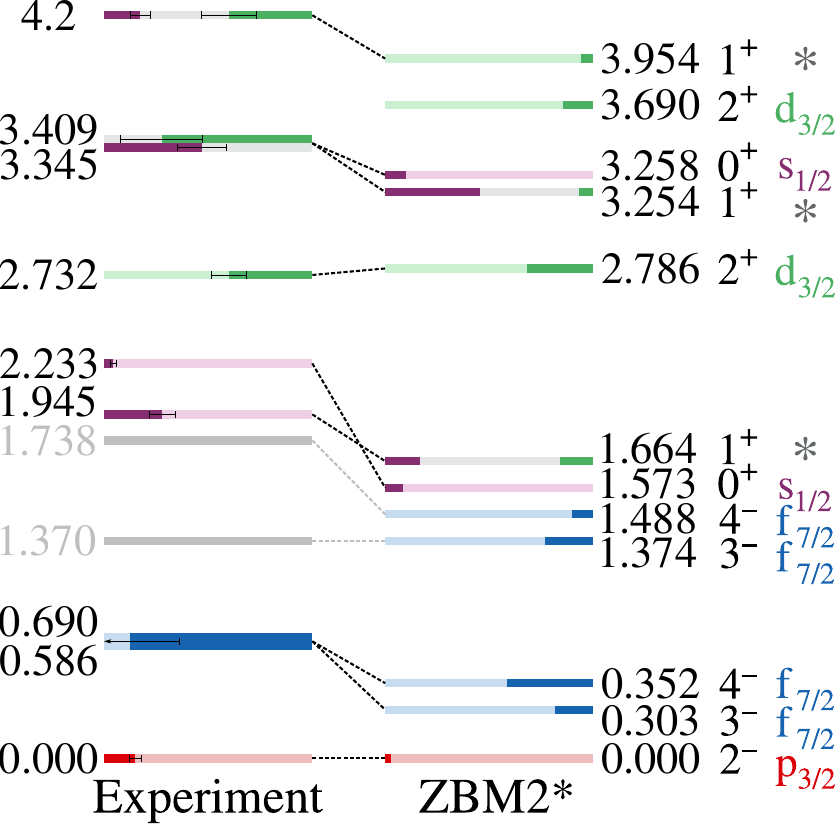}
     \caption{Comparison of $^{46}$K states and spectroscopic factors ($S$) resulting from this work (left) and from shell model calculations performed by B.\,A.~Brown~\cite{Note2} (right). Ths filled length of each state line represents $S$ as a fraction of the total possible $S$ for that orbital. The $s_{1/2}$ (purple) and $p_{3/2}$ (red) states fill from the left, and the $d_{3/2}$ (green) and $f_{7/2}$ (blue) states fill from the right. Mixed $s_{1/2}$$-d_{3/2}$ configuration states are marked with $*$.}
     \label{fig:46K_CS}
 \end{figure}

\begin{table}[]																			
\begin{tabular}{ccc|cc|cccccc}																			
\hline																			
\hline																			
\multicolumn{3}{c|}{This work}	&	\multicolumn{2}{c|}{Literature}	&	\multicolumn{5}{c}{Shell model}                                                       \\															
E$_{x}$	&	n$\ell_{j}$	&	S$_{sys}^{stat}$	&	E$_{x}$	&	$J^{\pi}$	&	E$_{x}$	&	$J^{\pi}$	&	n$\ell_{j}$	&	S	&	$\pi s_{1/2}^{-1}$	\\ \hline \hline
0.000	&	$p_{3/2}$	&	0.34$_{0.07}^{0.02}$	&	0.000	&	2$^{-}$	&	0.000	&	2$^{-}$	&	$p_{3/2}$	&	0.11	&	4.7\%	\\
0.586(3)	&	$f_{7/2}$	&	\ref{foot_1st}	&	0.587	&	3$^{-}$	&	0.304	&	3$^{-}$	&	$f_{7/2}$	&	1.47	&	35.8\%	\\
0.690(5)	&	$f_{7/2}$	&	\ref{foot_1st}	&	0.691	&	4$^{-}$	&	0.352	&	4$^{-}$	&	$f_{7/2}$	&	3.33	&	67.6\%	\\
---	&	---	&	---	&	1.370	&	3$^{-}$	&	1.374	&	3$^{-}$	&	$f_{7/2}$	&	1.85	&	42.5\%	\\
---	&	---	&	---	&	1.738	&	4$^{-}$	&	1.488	&	4$^{-}$	&	$f_{7/2}$	&	0.83	&	18.7\%	\\
1.945(3)	&	$s_{1/2}$	&	0.56$_{0.11}^{0.02}$	&	1.944	&	1$^{+}$	&	1.664	&	1$^{+}$	&	\multicolumn{2}{c}{\ref{foot_calc1MeV}}			&	61.2\%	\\
2.233(5)	&	$s_{1/2}$	&	0.10$_{0.02}^{0.01}$	&	2.222	&	0$^{+}$	&	1.573	&	0$^{+}$	&	$s_{1/2}$	&	0.18	&	30.5\%	\\
2.732(6)	&	$d_{3/2}$	&	1.6$_{0.3}^{0.2}$	&	2.790	&	2$^{+}$	&	2.768	&	2$^{+}$	&	$d_{3/2}$	&	1.27	&	43.2\%	\\
3.345(8)	&	\multicolumn{2}{c|}{\ref{foot_2nd}}			&	3.383	&	---	&	3.254	&	1$^{+}$	&	\multicolumn{2}{c}{\ref{foot_calc3MeV}}			&	67.4\%	\\
3.409(7)	&	\multicolumn{2}{c|}{\ref{foot_2nd}}			&	3.350	&	---	&	3.258	&	0$^{+}$	&	$s_{1/2}$	&	0.20	&	39.1\%	\\
---	&	---	&	---	&	---	&	---	&	3.690	&	2$^{+}$	&	$d_{3/2}$	&	0.58	&	21.3\%	\\
4.2(2)	&	\multicolumn{2}{c|}{\ref{foot_3rd}}			&	---	&	---	&	3.954	&	1$^{+}$	&	$d_{3/2}$	&	0.22	&	14.9\%	\\ \hline \hline
\end{tabular}																			
\footnotetext{\label{foot_1st}Unresolved in experiment, S($f_{7/2}$) = $7.0_{1.4}^{0.5}$}                      																			
\footnotetext{\label{foot_calc1MeV}Mixed in shell model only,  S($s_{1/2}$) = 0.34, S($d_{3/2}$) = 0.65}  																			
\footnotetext{\label{foot_2nd}Unresolved in experiment, S($s_{1/2}$) = $0.79_{0.16}^{0.08}$, S($d_{3/2}$) = $2.3_{0.5}^{0.7}$}   																			
\footnotetext{\label{foot_calc3MeV}Mixed in shell model only,  S($s_{1/2}$) = 0.91, S($d_{3/2}$) = 0.28}   																			
\footnotetext{\label{foot_3rd}Mixed in experiment only,  S($s_{1/2}$) = $0.19_{0.04}^{0.06}$, S($d_{3/2}$) = $1.5_{0.3}^{0.5}$} 																			
\caption{States in $^{46}$K observed in this work, compared to literature~\cite{NuclearDataSheets_Mass46} and ZBM2* calculations. Notably, the calculated percentage of $\pi s_{1/2}$-hole configuration in the wavefunction of each shell model state is indicated (see text).}																			
\label{table_46K_specfactors}																			
\end{table}

The comparison of the experimental results and shell model calculations for $^{46}$K are shown in Fig.~\ref{fig:46K_CS} and tabulated in Table~\ref{table_46K_specfactors}. In both cases, the only state that carries any $\nu p_{3/2}$ strength is the ground state, which is populated more strongly in the experiment than in the calculation. This discrepancy of a factor of three will be discussed further in Section~\ref{sect:protonFrac}.

The large total strength of $\nu f_{7/2}$ removal in $^{47}$K(d,t) is consistent with the very small $\nu f_{7/2}$ addition strength revealed by the 0.728~MeV state in $^{47}$K(d,p). The shell model expects the neutron-removal strength to be shared between four states -- 18\% in $3_1^-$, 42\% in $4_1^-$, 23\% in $3_2^-$ and 10\% in $4_2^-$. Experimentally, we find a concentration of strength,$\sim90\%$ , in the $3_1^-$$-4_1^-$ pair, with an uncertain population of the high-energy pair. As we cannot say with certainty that the high-energy pair are populated, due to the inconsistent $\gamma$-decay observations, we here simply report an agreement with the $4_1^-$ being the strongest $\nu f_{7/2}$ state. We do wish to highlight, however, that the shell model predicts similar neutron structures for the four states, with the key variance being the proton configurations; that is, the $4_1^-$ state is 68\% $\pi(s_{1/2}^{-1})$, whereas the $4_2^-$ is 70\% $\pi(d_{3/2}^{-1})$. This suggests that the proton configuration mixing between these states may be poorly estimated in the shell model -- this is particularly interesting as, from a simple assessment of the known structure of $^{46}$K, shown in Fig.~\ref{fig:46K_levelscheme}, one would expect that the low-energy $2^-$$-3^-$$-4^-$$-5^-$ quadruplet would originate from $\pi(d_{3/2}^{-1})$$\nu(f_{7/2}^{-1})$, and the higher-energy $3^-$$-4^-$ pair to originate from $\pi(s_{1/2}^{-1})$$\nu(f_{7/2}^{-1})$. Our experimental results, and the shell model calculations, suggest that the $4^-$ states have mixed to such an extent that the $4^-$ state structures are nearly inverted.

The distinct shape of the $\nu s_{1/2}$ differential cross section and the comparatively high statistics of the $s_{1/2}$-carrying states leads to greater confidence in these spectroscopic factors, in comparison to other states. Indeed, the observed $\nu s_{1/2}$ spectroscopic factors are relatively well-predicted by the shell model, and the sum of spectroscopic factors for theory and experiment are in agreement, with $\sum$S$_{exp}$ = 1.6 $\pm$ 0.2, and $\sum$S$_{calc}$ = 1.63. This supports the natural assumption of an occupied $\nu s_{1/2}$ orbital. Interestingly, the shell model expects the $1^+_1$ state to be mixed $s_{1/2}$-$d_{3/2}$, but the experimental differential cross section is consistent with a pure $s_{1/2}$ configuration.

The sum of the observed $\nu d_{3/2}$ strength is complicated by the fact that only one pure-$d_{3/2}$ state is observed, and the remaining strength is all mixed with strong $\nu s_{1/2} $ states. As the shape of the $d_{3/2}$ differential cross-section is relatively flat in the angular ranges covered here, measurements of the mixed-configuration states are unable to place tight constraints on the magnitude of the $d_{3/2}$ contribution. As such, the sum of $d_{3/2}$ strength -- $\sum S_{calc}=5.4\pm2.5$ -- is large compared to the shell model calculation of $\sum S_{calc}=3.00$, but is also very imprecise. This imprecision hampers comparison to theory somewhat, but some conclusions can still be drawn; notably, of the two expected pure-$\nu(d_{3/2})$ $2^+$ states found in the calculation, the higher-energy state has no experimental analogue. Interestingly, this echoes the earlier discussion of $\nu f_{7/2}$ states; the state we observe is $\pi s_{1/2}^{-1}$-dominant, and the state that we do not observe is predicted by the shell model to be $\pi d_{3/2}^{-1}$-dominant (see Table~\ref{table_46K_specfactors}), perhaps revealing a poor estimation of the level of $\pi s_{1/2}^{-1}$ mixing in the unobserved state.

Finally, we note that the $s_{1/2}$ strength generally appears at lower excitation energies than the $d_{3/2}$ strength, in both the experimental data and the shell model calculations. This is contrary to what would be naïvely expected from the "normal" ordering of states, wherein the $\nu s_{1/2}$ should be deeper (and hence more bound) than the $\nu d_{3/2}$. As such, our experimental data supports a reordering of these neutron orbitals, which is well-reflected in the shell model. This is further supported by single neutron-removal studies of $^{48}$Ca~\cite{Ota2025_NDS_Neq47,martin1972_48Ca-pd}, which find that two states close in energy, at 2.58~MeV and 2.60~MeV, deplete more than 90\% of the $\nu d_{3/2}$ and $\nu s_{1/2}$ strength, respectively, indicating a near-degeneracy of these neutron orbitals in the region of $^{48}$Ca.

\subsection{Systematic analysis regarding $^{46}$Ar}

Here, we present a slightly extended discussion of a result which was noted briefly in Ref.~\cite{Paxman2025_47Kdp}, which can now be expanded following the recent work of Brugnara~\textit{et al.}~\cite{Brugnara2025_arXiv}. In that work, the authors present a study of $^{46}$Ar(${^3}$He,d)$^{47}$K, concluding that the ground state of $^{46}$Ar has a fully depleted $\pi s_{1/2}$ orbital, as opposed to the previous assumption of equiprobable population of the $\pi(s_{1/2},d_{3/2})$ orbitals~\cite{Gaudefroy2006_46Ardp47Ar,Gaudefroy2006_46Ardp_SignoracciComment,Gaudefroy2006_46Ardp_Reply}. This further contextualizes the comparison noted in Ref.~\cite{Paxman2025_47Kdp}, where the single-particle energies of the $\nu f_{5/2}$, $p_{1/2}$ and $p_{3/2}$ orbitals populated by $^{47}$K(d,p) were found to lie directly between those observed in $^{46}$Ar(d,p) and $^{48}$Ca(d,p). Recalling the $\pi(s_{1/2}^2,d_{3/2}^4)$ structure of $^{48}$Ca$_{g.s.}$, and the $\pi(s_{1/2}^1,d_{3/2}^4)$ structure of $^{47}$K$_{g.s.}$, then this smooth variation in the neutron single particle energies would seem to support the closed-shell $\pi(s_{1/2}^0,d_{3/2}^4)$ structure of $^{46}$Ar$_{g.s.}$ concluded by Brugnara \textit{et al.}, from which they propose a possible proton ``bubble''~\cite{Mutschler2017_34Si_BubbleNucleus,Saxena2019_BubbleNUcleiTheoryStudy} and a new proton magic number at $Z=18$ in the region of $N=28$.

\subsection{Proton configuration\label{sect:protonFrac}}

Several results presented in this paper have indicated that there may be some imprecision in the shell model predictions of proton configuration mixing, particularly in relation to the strength with which states are populated. Recall, for example, the non-observation of some predicted states in $^{47}$K(d,t), namely the higher-energy 3$^-$/4$^-$ $f_{7/2}$ pair and the highest-energy 2$^+$ $d_{3/2}$ state. Additionally, it was noted in Ref.~\cite{Paxman2025_47Kdp} that there is a significant disparity between the experimental $^{47}$K(d,p) spectroscopic factors and those predicted by the shell model, seemingly also related to the proton configuration. We here explore this disparity further, comparing the accuracy of the shell model spectroscopic factors to the calculated proton structure. In this case, we define the accuracy as the difference between the calculated spectroscopic factor, $S_{calc}$, and the experimental spectroscopic factor, $S_{meas}$, scaled by the maximum possible spectroscopic factor, $S_{max}$, to allow comparison between (d,p) and (d,t) transfer. For this analysis, some states have been excluded; namely those without clear shell model analogues, with mixed configurations, inextricable doublets or unbound states. After these exclusions, all states in $^{48}$K up to 3.9~MeV are included, except the unpopulated 0.279~MeV state. Unfortunately, only three $^{46}$K data points are suitable for this analysis -- the ground state, the 2.233 MeV, and the 2.732 MeV states -- of which the latter has a large error bar, resulting from the flatness of the $\nu d_{3/2}$ differential cross section.

\begin{figure}
    \centering
    \includegraphics[width=\linewidth]{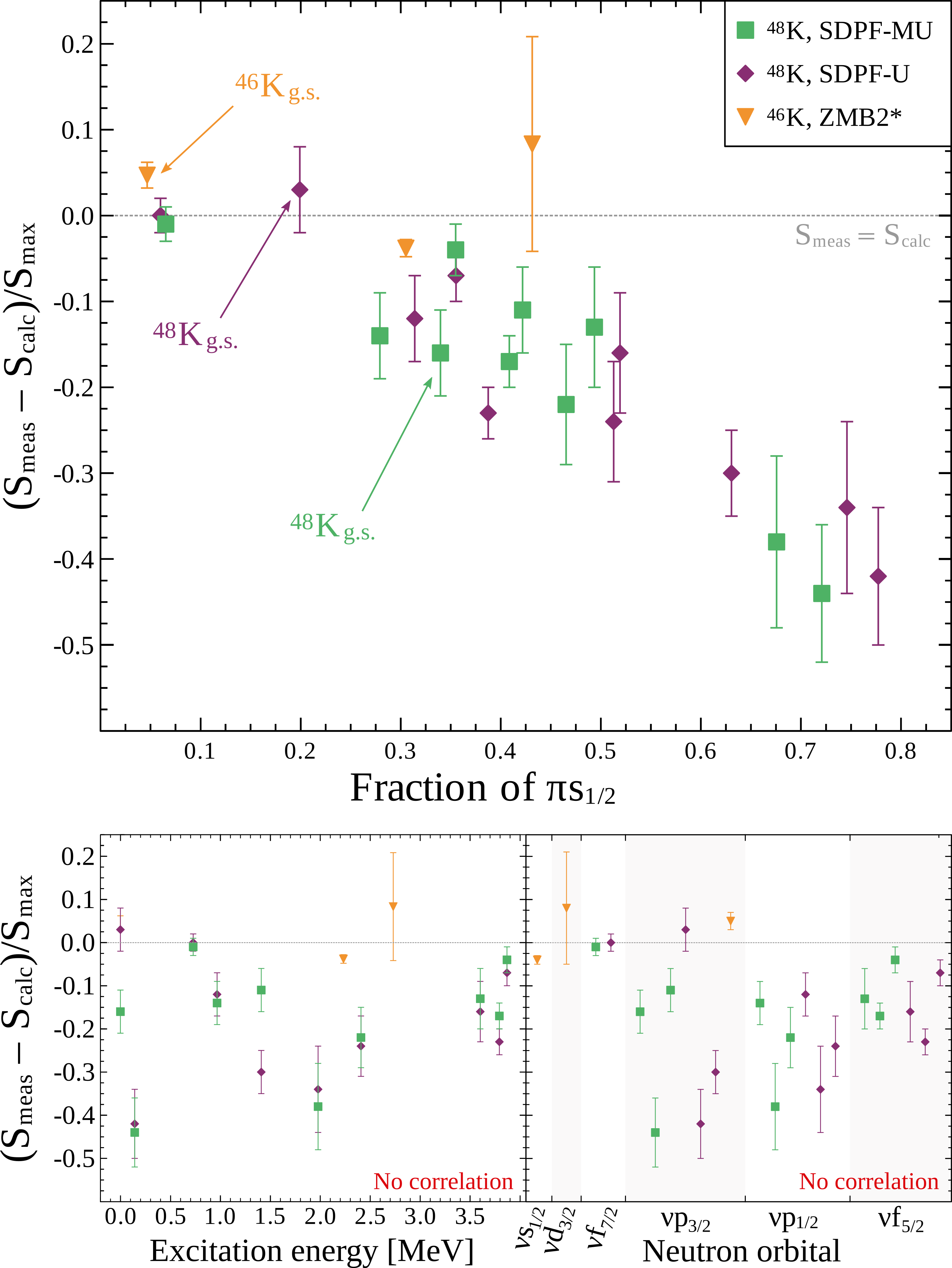}
    \caption{(Top) Accuracy of the three shell model calculations as a function of the calculated $\pi s_{1/2}^{-1}$ component of the state wavefunction, showing an apparent correlation. See text for details. (Bottom) Accuracy of the shell model calculations, plotted firstly against the excitation energy of the state and secondly against the transferred neutron orbital. Notably, there is no apparent correlation between these variables. }
    \label{fig:ProtonFrac}
\end{figure}

An apparent pattern has emerged in this analysis -- shown in Fig.~\ref{fig:ProtonFrac} -- wherein the states with large predicted $\pi s_{1/2}^{-1}$ components are systematically overestimated in the calculation, whereas states with smaller $\pi s_{1/2}^{-1}$ components are well-matched or underestimated. The ground states are a convenient illustration of this point; the SDPF-U predicts a $\pi s_{1/2}^{-1}$ component half the size of SDPF-MU, and indeed, SDPF-U is also much more accurate in it's prediction of the ground state spectroscopic factor than SDPF-MU. Further, the ground state of $^{46}$K has a very small calculated $\pi s_{1/2}^{-1}$ component of 4.7\%, making it a cornerstone measurement in the low-$\pi s_{1/2}^{-1}$ region. Indeed, this state is underestimated by a factor of three in the shell model calculations. The only other data points at very low $\pi s_{1/2}^{-1}$ are the two calculations of the 0.728~MeV state in $^{48}$K, which are found to be within error of experimental observation. Critically, this apparent relationship holds irrespective of the $\ell$-transfer of the neutron, the spin-orbit splitting, the excitation energy of the final state, or whether the neutron is being added or removed. Hence, the shell model's inaccuracy of the neutron transfer spectroscopic factors seems to be independent of the neutron configuration, and depends primarily on the proton configuration. This correlation provides a possible point of comparison for future shell model interactions, benchmarking their ability to effectively capture the complex proton structure which is so critical to understanding the single-particle origins of the $N=28$ island of inversion.

\section{Summary}

In this paper, we present the first measurement of the $^{47}$K(d,t$\gamma$)$^{46}$K direct transfer reaction, alongside additional details of the concurrent $^{47}$K(d,p$\gamma$) analysis~\cite{Paxman2025_47Kdp}. By exploiting the exotic \piSOneDFour{} proton structure of $^{47}$K$_{g.s.}$ -- as opposed to the more typical \piSTwoDThree{} structures of $^{46}$K$_{g.s.}$ and $^{48}$K$_{g.s.}$ -- the combination of these neutron-removal and neutron-addition analyses amounts to a detailed scan of the interaction between one specific proton orbital, $\pi s_{1/2}$, and a broad range of neutron orbitals. These span the whole $fp$-shell and most of the $sd$-shell, encompassing the canonical $N=20$ and $N=28$ magic numbers. Through comparison to leading shell-model calculations, we find a good replication of the energies and spin-parities of observed states in both nuclei, but a significant disagreement in spectroscopic factors. This is traced to an apparent failure of the shell model to correctly capture the complex proton behavior in this region of near-degeneracy between $\pi s_{1/2}$ and $\pi d_{3/2}$, exemplified most strikingly in Fig.~\ref{fig:ProtonFrac}, where a clear relationship emerges, indicating that the proton configuration mixing is the key limitation in the prediction of neutron-transfer spectroscopic factors. As the lowest-$\pi s_{1/2}^{-1}$ fraction state to be simply interpreted in this work, the $^{46}$K ground state works to anchor this relationship in the neutron-removal case. As such, we present these results as a benchmark for interactions between $\pi s_{1/2}$ and neutron orbitals above and below the $N=28$ shell gap, which we expect to be particularly important for a microscopic understanding of the $N=28$ island of inversion, centered around $^{42}$Si ($\pi s_{1/2}^0$) and $^{44}$S ($\pi s_{1/2}^2$).


\begin{acknowledgments}
We acknowledge the GANIL facility for provision of heavy-ion beams, and we thank J. Goupil, G. Fremont, L. M{\'e}nager, and A. Giret for assistance in using the G1 beam line and its instrumentation. We acknowledge
the AGATA collaboration for the use of the spectrometer. This work was supported by the Science Technology Facilities Council (United Kingdom) under the grants ST/N002636/1, ST/P005314/1 and ST/V001108/1. This work was also partially funded by MICIU MCIN/AEI/10.13039/501100011033, Spain with grants PID2020-118265GB-C42, PRTR-C17.I01, Generalitat Valenciana, Spain with grant CIPROM/2022/54, ASFAE/2022/031, CIAPOS/2021/114 and by the EU NextGenerationEU funds. 
Additional funcing from Spanish grant PID2021-127711NB-100. This work was partially supported by the U.S. Department of Energy, Office of Science, Office of Nuclear Physics, under contract number DE-AC02-06CH11357. 
\end{acknowledgments}

\bibliographystyle{apsrev4-1}
\bibliography{Other/biblio}

  
\end{document}